\documentclass[pdflatex,sn-standardnature]{sn-jnl}% Default with double column layout
\usepackage{array}
\newcolumntype{C}{>{$}c<{$}} % math-mode version of "c" column type

\begin{document}

%Length limit
%First paragraph: 200 words
%Total main text including first paragraph: 2500 words
%4 display items that each occupy a quarter of a page
%50 references
%Sections are separated with subheadings to aid navigation. Subheadings may be up to 40 characters (including spaces).

%\title[GJ\,1214b Phase Curve]{A thermal emission phase curve of the sub-Neptune exoplanet GJ\,1214b}

\title[GJ\,1214b Phase Curve]{A reflective, metal-rich atmosphere for GJ\,1214b from its JWST phase curve}

\author[1]{Eliza M.-R. Kempton} %\email{ekempton@astro.umd.edu}

\author[2]{Michael Zhang}
\author[2]{Jacob L.\ Bean}
\author[3]{Maria E.\ Steinrueck}
\author[4]{Anjali A.\,A.\ Piette}
\author[5,6]{Vivien Parmentier}
\author[7]{Isaac Malsky}
\author[8]{Michael T.\ Roman}
\author[7]{Emily Rauscher}
\author[4]{Peter Gao}
\author[9]{Taylor J.\ Bell}
\author[2]{Qiao Xue}
\author[5,10]{Jake Taylor}
\author[1,11]{Arjun B.\ Savel}
\author[1]{Kenneth E.\ Arnold}
\author[1]{Matthew C.\ Nixon}
\author[12]{Kevin B.\ Stevenson}
\author[13]{Megan Mansfield}
\author[14]{Sarah Kendrew}
\author[3,15]{Sebastian Zieba}
\author[16,17]{Elsa Ducrot}
\author[17]{Achr\`{e}ne Dyrek}
\author[17]{Pierre-Olivier Lagage}
\author[18]{Keivan G. Stassun}
\author[19]{Gregory W.\ Henry}
\author[20]{Travis Barman}
\author[21]{Roxana Lupu}
\author[1]{Matej Malik}
\author[22]{Tiffany Kataria\textsuperscript{22}}
\author[1]{Jegug Ih}
\author[23]{Guangwei Fu}
\author[24]{Luis Welbanks}
\author[25]{Peter McGill}

\affil[1]{Department of Astronomy, University of Maryland, College Park, MD, USA}
\affil[2]{Department of Astronomy \& Astrophysics, University of Chicago, Chicago, IL, USA}
\affil[3]{Max-Planck-Institut f\"ur Astronomie, Heidelberg, Germany}
\affil[4]{Earth and Planets Laboratory, Carnegie Institution for Science, Washington, DC, USA}
\affil[5]{Department of Physics, University of Oxford, Oxford, UK}
\affil[6]{Université Côte d’Azur, Observatoire de la Côte d’Azur, CNRS, Laboratoire Lagrange, France}
\affil[7]{Department of Astronomy, University of Michigan, Ann Arbor, MI, USA}
\affil[8]{School of Physics and Astronomy, University of Leicester, Leicester, UK}
\affil[9]{BAER Institute, NASA Ames Research Center, Moffet Field, CA, USA}
\affil[10]{Institut Trottier de Recherche sur les Exoplan\`etes and D\'epartement de Physique, Universit\'e de Montr\'eal, Montr\'eal, QC, Canada}
\affil[11]{Center for Computational Astrophysics, Flatiron Institute, New York, NY, USA}
\affil[12]{Johns Hopkins Applied Physics Laboratory, Laurel,  MD, USA}
\affil[13]{Steward Observatory, University of Arizona, Tucson,  AZ, USA}
\affil[14]{European Space Agency, Space Telescope Science Institute, Baltimore, MD, USA}
\affil[15]{Leiden Observatory, Leiden University, Leiden, The Netherlands}
\affil[16]{Paris Region Fellow, Marie Sklodowska-Curie Action}
\affil[17]{AIM, CEA, CNRS, Universit\'e Paris-Saclay, Universit\'e de Paris, Gif-sur-Yvette, France}
\affil[18]{Department of Physics and Astronomy, Vanderbilt University, Nashville, TN, USA}
\affil[19]{Center of Excellence in Information Systems, Tennessee State University, Nashville, TN, USA}
\affil[20]{Lunar and Planetary Laboratory, University of Arizona, Tucson, AZ, USA}
\affil[21]{Eureka Scientific, Inc., Oakland, CA, USA}
\affil[22]{NASA Jet Propulsion Laboratory, California Institute of Technology, Pasadena,  CA, USA}
\affil[23]{Department of Physics \& Astronomy, Johns Hopkins University, Baltimore, MD, USA}
\affil[24]{School of Earth \& Space Exploration, Arizona State University, Tempe, AZ, USA}
\affil[25]{Department of Astronomy \& Astrophysics, University of California, Santa Cruz, CA, USA}

\maketitle

\textbf{There are no planets intermediate in size between Earth and Neptune in our Solar System, yet these objects are found around a substantial fraction of other stars \cite{howard12}. Population statistics show that close-in planets in this size range bifurcate into two classes based on their radii \cite{fulton17,vaneylen18}. It is hypothesized that the group with larger radii (referred to as ``sub-Neptunes'') is distinguished by having hydrogen-dominated atmospheres that are a few percent of the total mass of the planets \cite{bean21}. GJ\,1214b is an archetype sub-Neptune that has been observed extensively using transmission spectroscopy to test this hypothesis \cite{bean10, croll11, bean11, desert11, berta12, fraine13, kreidberg14, kasper20, orell22, spake22}. However, the measured spectra are featureless, and thus inconclusive, due to the presence of high-altitude aerosols in the planet’s atmosphere. Here we report a spectroscopic thermal phase curve of GJ\,1214b obtained with JWST in the mid-infrared. The dayside and nightside spectra (average brightness temperatures of 553\,$\pm$\,9 and 437 $\pm$\,19\,K, respectively) each show $>3\sigma$ evidence of absorption features, with H$_2$O as the most likely cause in both. The measured global thermal emission implies that GJ\,1214b's Bond albedo is 0.51\,$\pm$\,0.06. Comparison between the spectroscopic phase curve data and three-dimensional models of GJ\,1214b reveal a planet with a high metallicity atmosphere blanketed by a thick and highly reflective layer of clouds or haze.}

% There are no planets intermediate in size between Earth and Neptune in our Solar System, yet these objects are found around a substantial fraction of other stars. Population statistics show that close-in planets in this size range bifurcate into two classes based on their radii. It is hypothesized that the group with larger radii (referred to as ``sub-Neptunes'') is distinguished by having hydrogen-dominated atmospheres that are a few percent of the total mass of the planets. The transiting exoplanet GJ 1214b is an archetype sub-Neptune that has been observed extensively using transmission spectroscopy to test this hypothesis. However, the results have been inconclusive because high-altitude aerosols have prevented the measurement of the planet's atmospheric composition. To circumvent the challenges of featureless transmission spectra, here we present a spectroscopic thermal phase curve of GJ 1214b with the JWST MIRI instrument. The phase curve data reveal that the planet's dayside spectrum resembles a blackbody at 553 +/- 5 K, whereas its nightside spectrum  is not well fit by a single blackbody and reveals spectral features suggestive of H2O and perhaps CH4 and HCN.  The planet's global thermal emission allows us to measure GJ 1214b's Bond albedo to be 0.51 +/- 0.04, implying reflective aerosols. Comparison between our spectroscopic phase curve data and three-dimensional models of GJ 1214b reveal a planet with a high metallicity atmosphere blanketed by a thick and highly reflective layer of clouds or haze.

The exoplanet GJ\,1214b has a radius of 2.6\,$R_{\oplus}$ and orbits its late M dwarf host star with a period of 37.9\,hours \cite{charbonneau09}. We observed the phase curve of GJ\,1214b using JWST's Mid-Infrared Instrument Low Resolution Spectrometer (MIRI LRS) \cite{kendrew15} on July 20\,--\,22, 2022. The observation was a time series of regular and continuous integrations using the slitless prism mode, starting 2.0~hours before the predicted time of secondary eclipse.  The data acquisition continued through the eclipse, a transit, and for 1.1 hours after a second eclipse for a total of 41.0\,hours. The telescope pointing was kept fixed during the observation; neither scanning nor dithering was used. A total of 21,600 integrations with 42 groups per integration (6.68\,s integration time) were obtained. 

\begin{figure}[t]
    \centering
    \includegraphics[width=1.0\textwidth]{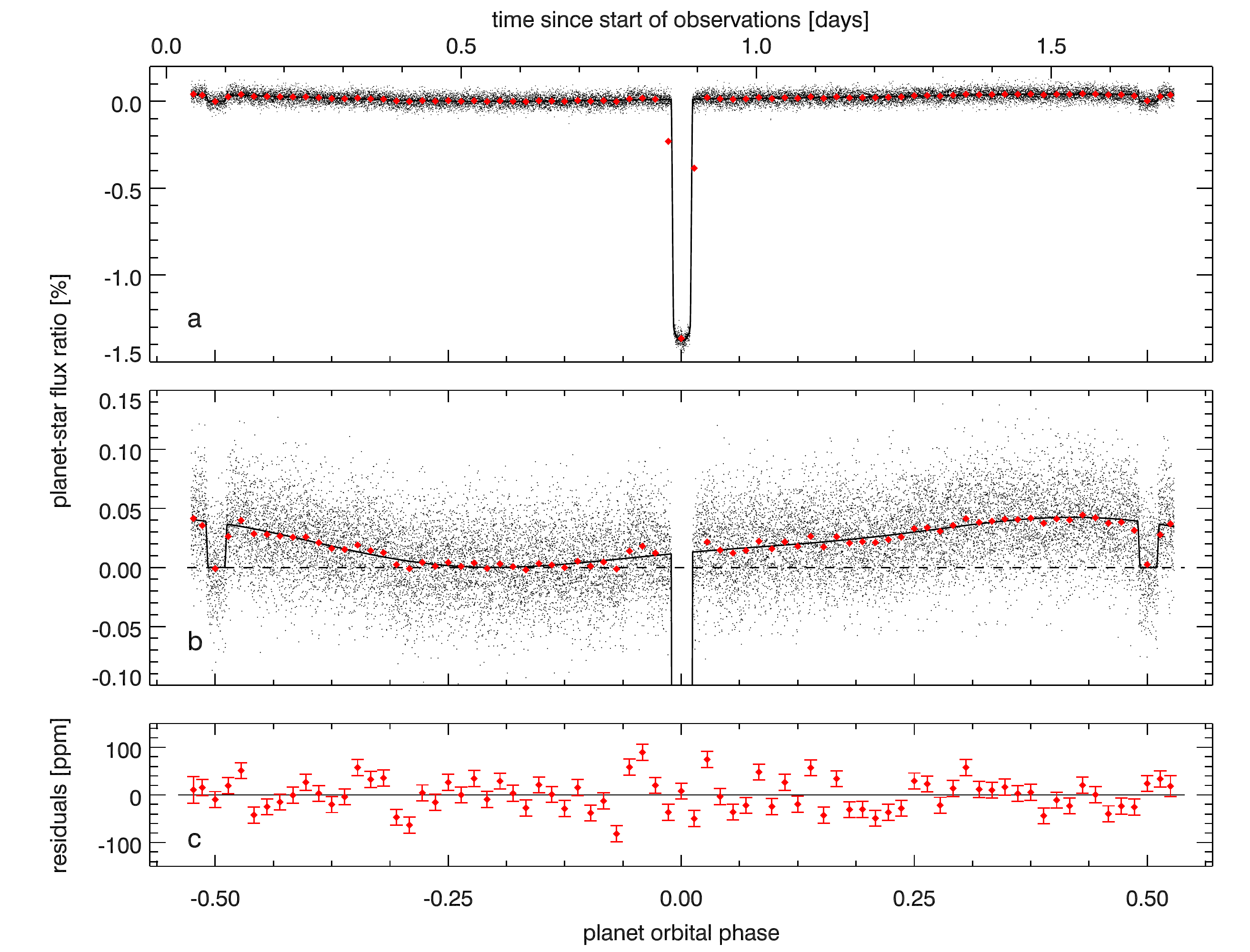}
    \caption{\textbf{The white light phase curve of GJ\,1214b.} \textbf{a,} The phase curve integrated from $5-12$\,$\mu$m after subtraction of instrument systematics and removal of the first hour of data.  The transit and two eclipses are clearly seen at phase 0.0 and $\pm 0.5$, respectively.  Red points are binned at every 5 degrees in orbital phase.  The black line is our best-fit astrophysical model, which includes the primary transit, secondary eclipses, and phase-dependent thermal emission assuming a second-order sinusoid functional form.  \textbf{b,} Same as panel a, but zoomed in to show the phase modulation in the planet's thermal emission.  The dashed black line indicates the (presumed constant) stellar flux in the absence of any emission from the planet.  \textbf{c,} Residuals of the binned data from the astrophysical model with 1$\sigma$ error bars. }
    \label{fig:wl_phasecurve} 
\end{figure}

We used a custom pipeline to reduce the data and extract the combined spectra of the planet and its host star from 5 to 12\,$\mu$m. We generated spectroscopic light curves (Extended Data Figure~\ref{fig:spectroscopic_lc}, inverted to equivalently produce spectra at each orbital phase; Extended Data Figure~\ref{fig:spectra_vs_phase}) by binning the data by 0.5\,$\mu$m (corresponding to 7 to 28 pixels per bin). We also produced a band-integrated ``white'' light phase curve by summing the data over all wavelengths (Figure~\ref{fig:wl_phasecurve}). Although the raw JWST light curves exhibit systematics that are typical for space-based phase curve observations, we clearly see the transit and the secondary eclipse in the light curve prior to any detrending (Extended Data Figure~\ref{fig:raw_white_lc}). We thus applied corrections for the systematics using standard methods and fit the data with an exoplanet phase curve model. More details of the data analysis are given in the Methods section.

Previous observations at 4.5\,$\mu$m with the Spitzer Space Telescope tentatively detected the secondary eclipse of GJ\,1214b half an orbital period following the transit (i.e., at phase 0.5) with a corresponding brightness temperature of $545^{+40}_{-50}$\,K \cite{gillon14}. We confirm the timing of the secondary eclipse, which is also consistent with prior constraints from radial velocity observations \cite{cloutier21}, suggesting a nearly circular orbit for the planet.  We measure a best-fit brightness temperature for the MIRI 5\,--\,12\,$\mu$m secondary eclipse of $553 \pm 9$\,K, in further agreement with the prior Spitzer observation.

\begin{figure}[h]
    \centering
    \includegraphics[width=1.0\textwidth]{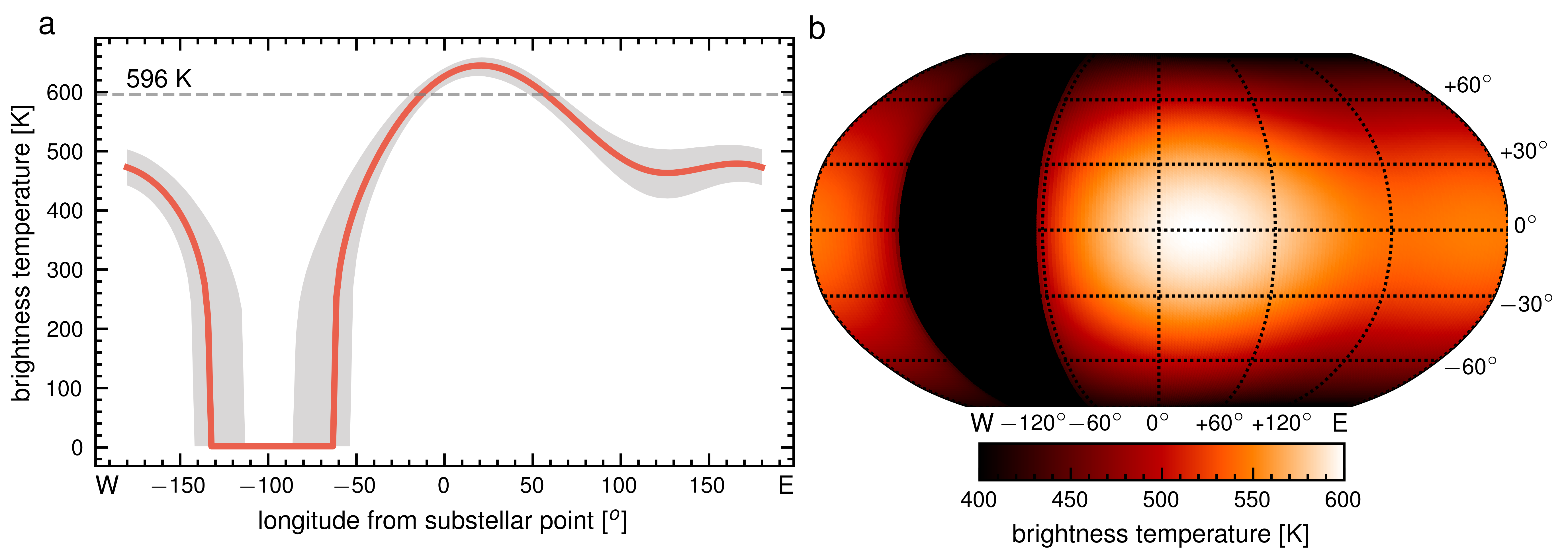}
    \caption{\textbf{Temperature map of GJ 1214b.} \textbf{a,} The equatorial 5\,--\,12\,$\mu$m brightness temperature vs.\ phase angle of GJ\,1214b, obtained by inverting the phase curve observations, as described in the Methods section.  The gray region is the 3$\sigma$ confidence interval on the derived temperature.  The dashed line indicates the zero-albedo temperature of GJ\,1214b under conditions of uniform heat redistribution.  \textbf{b,} The photospheric temperature map extrapolated by inverting the phase curve assuming a cosine dependence of the temperature with latitude.  Black regions are where the mapping inversion of the measured planet-star flux ratio produces negative planetary emission due to the functional form that is enforced \cite{keating17}.}
    \label{fig:T_map} 
\end{figure}

Our measurement of GJ\,1214b's thermal emission allows us to map out the planet's longitudinal brightness temperature distribution in the 5\,--\,12~$\mu$m wavelength range (Figure~\ref{fig:T_map}).  It is apparent from this calculation that the planet must have a non-zero albedo, as its emission at nearly all longitudes falls well below what is expected for a fully absorptive planet in the limit that it uniformly redistributes the energy received from its host star (dashed line in Figure~\ref{fig:T_map}).  Furthermore, we estimate that the MIRI LRS bandpass encompasses approximately 50\,--\,60\% of the planet's emitted flux (see Methods).  This gives us confidence that we are capturing the majority (and the peak) of GJ\,1214b's thermal emission and allows us to determine the planet's Bond albedo without heavy reliance on model extrapolations.  We estimate a Bond albedo of \mbox{$0.51 \pm 0.06$}, implying that the planet reflects a considerable fraction of the incident starlight it receives.  The error bar is derived formally from the phase curve data;  we estimate that systematic uncertainty in the nightside flux from choices in the data reduction could make the error as large as 0.12, as detailed in the Methods.  For context, hot Jupiter exoplanets have been found to have very low geometric and Bond albedos \cite{rowe08, stevenson14, brandeker22}.  The majority of solar system planets have Bond albedos less than 0.35, with notable exceptions being Venus (0.75) \cite{moroz81} and Jupiter (new upwardly revised value of 0.53) \cite{li18}.

%we estimate that systematic uncertainty in the nightside flux from choices in the data reduction could make the error as large as 0.12.

%We estimate a Bond albedo of $0.51 \pm 0.07$, implying that the planet reflects a considerable fraction of the incident starlight it receives (see details in the Methods section).   For context, hot Jupiter exoplanets have been found to have very low geometric and Bond albedos \cite{rowe08, stevenson14, brandeker22}.  The majority of solar system planets have Bond albedos less than 0.35, with notable exceptions being Venus (0.75) \cite{moroz81} and Jupiter (new upwardly revised value of 0.53) \cite{li18}.  

\begin{figure}[t]
    \centering
    \includegraphics[width=1\textwidth]{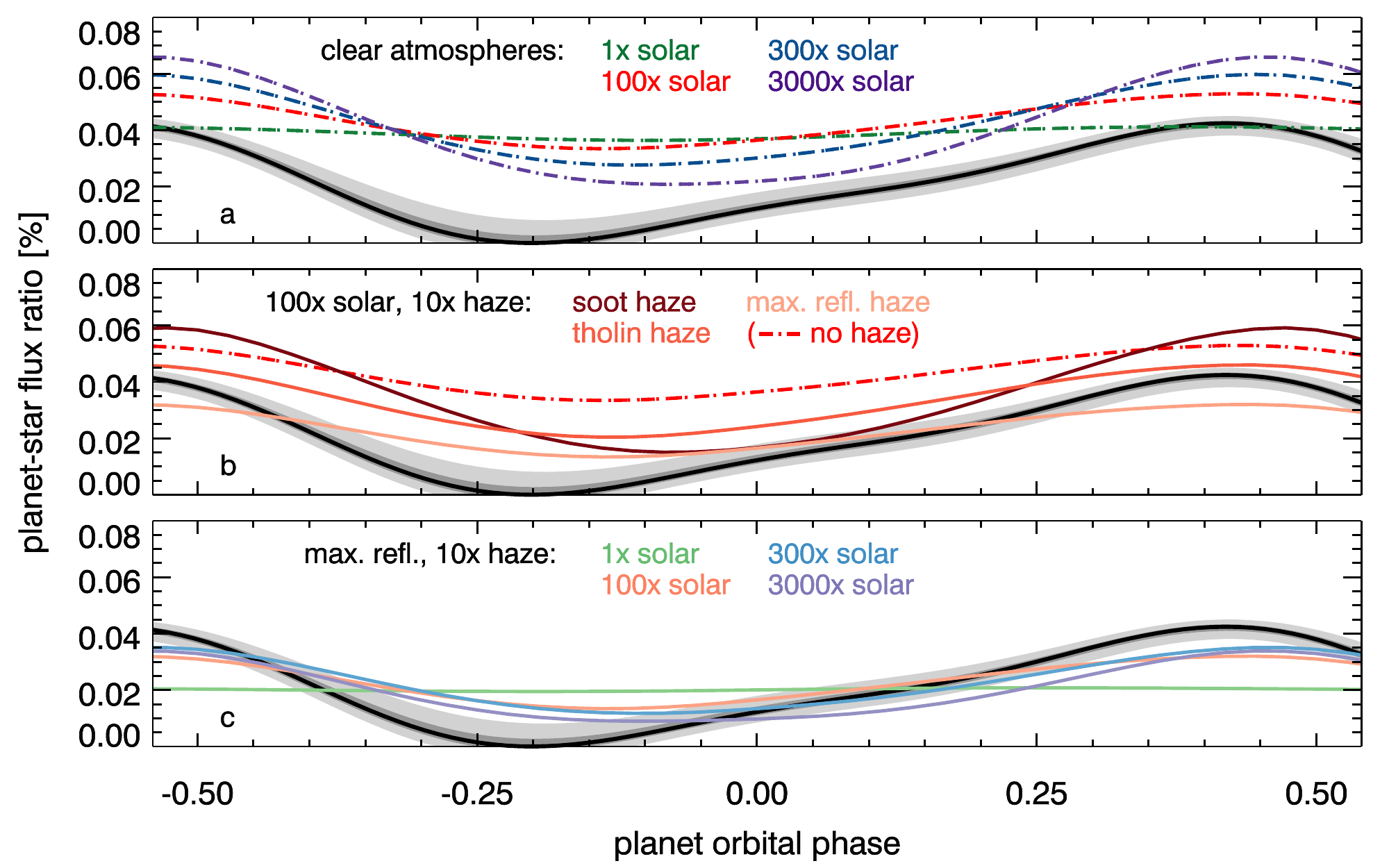}
    \caption{\textbf{White light phase curve compared to GCM outputs.} \textbf{a,} The best-fit phase curve model (thick black line) is compared against GCM outputs for clear-atmosphere models, assuming various metallicities, as indicated.  Dark and light gray shaded regions are the 1$\sigma$ and 3$\sigma$ uncertainty regions, respectively. All clear atmosphere models predict the planet to be much hotter and therefore have considerably more 5 -- 12\,$\mu$m thermal emission than what is observed.  \textbf{b,} The addition of a thick haze to the 100$\times$ solar metallicity GCM (solid colored lines) alters the model predictions.  Absorptive hazes (e.g., soot) heat the dayside and cool the nightside, producing even stronger dayside emission and thus a poor fit to the observations.  More reflective hazes (e.g., tholins and ``maximally reflective" hazes with a high imposed SSA) cool the planet globally and provide a better match to the observed dayside flux.  \textbf{c,} Models with reflective hazes and high metallicity (e.g., blue and purple lines) provide the best qualitative fit to the observed phase curve, although further tuning of the haze parameters and metallicity would be required to reproduce the observed thermal emission at all orbital phases.}
    \label{fig:wl_models} 
\end{figure}

\begin{figure}[t]
    \centering
    \includegraphics[width=1.0\textwidth]{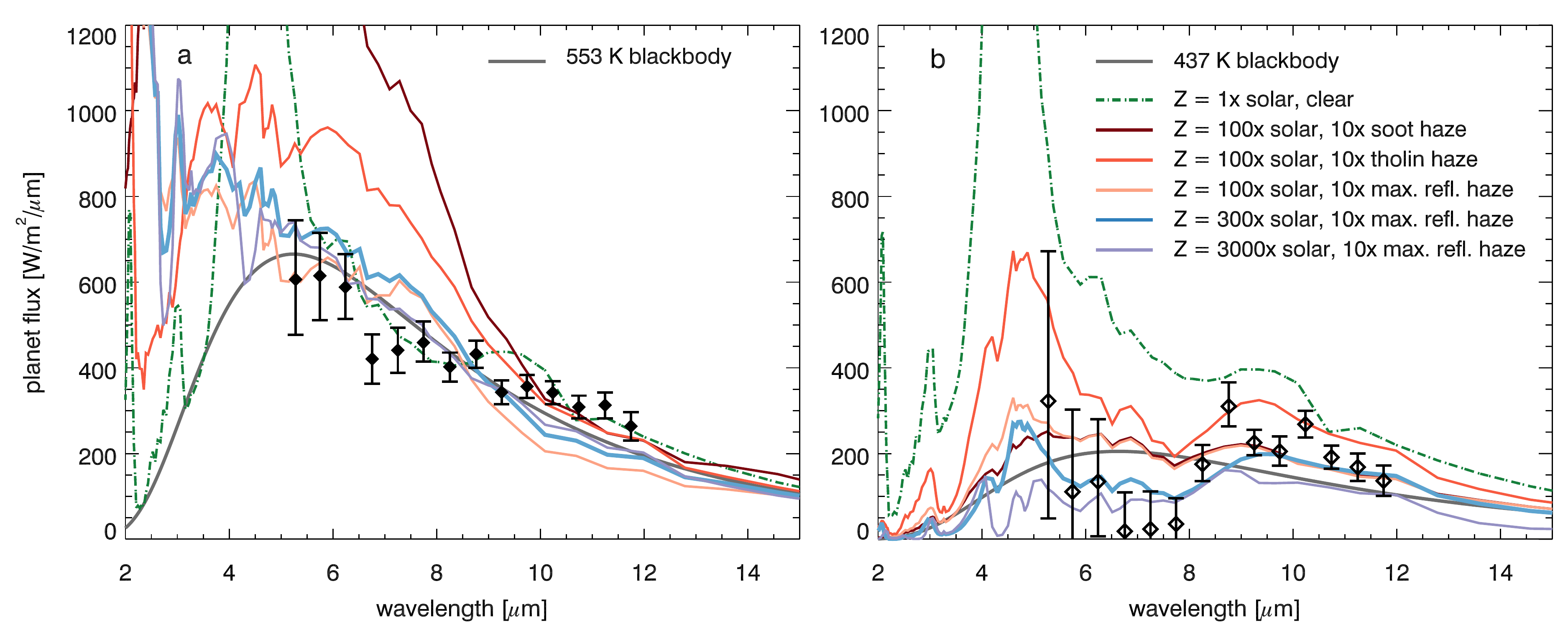}
    \caption{\textbf{JWST MIRI dayside and nightside spectra of GJ\,1214b.} \textbf{a,} The dayside spectrum is shown in black (filled symbols with 1$\sigma$ error bars) with the best-fit blackbody curve at $T = 553$\,K overplotted in gray.  \textbf{b,} The nightside spectrum (open symbols with 1$\sigma$ error bars) and its best-fit blackbody of $T = 437$\,K.  In both panels, colored lines show spectra derived from GCMs with various metallicities (1 -- 3000$\times$ solar), haze optical thickness (clear atmospheres and 10$\times$ thicker than our nominal haze model), and haze optical properties (soot, tholin, and highly reflective), as described in the Methods section.  The dayside and nightside spectra are jointly well-fit by models that have high metallicity accompanied by thick and reflective haze (e.g., the solid light blue lines, representing a hazy atmosphere at 300$\times$ solar metallicity), suggesting a globally consistent solution.}
    \label{fig:fp_day_night} 
\end{figure}

To obtain further constraints on GJ\,1214b's atmospheric composition, aerosol properties, and atmospheric dynamics, we ran a new set of 3-D general circulation models (GCMs) spanning compositions from solar metallicity to high mean molecular weight atmospheres (i.e., 3000$\times$ solar metallicity; see Methods).  Transmission spectroscopy of GJ\,1214b requires a thick aerosol layer at the planet's terminator \cite{kreidberg14}.  The composition of the aerosols is unknown, but hydrocarbon haze is the favored culprit \cite{morley15,kawashima19b,adams19,lavvas19,gao20b}.   Prior 3-D modeling of GJ\,1214b has focused on clear atmospheres \citep{kataria14} and condensate clouds \cite{charnay15b,charnay15,christie22} but neglected photochemical hazes. Given large uncertainties in the nature of GJ\,1214b's aerosols, we ran end-member GCMs both with clear atmosphere conditions and with a thick global haze.  Our nominal haze model uses the optical properties of soot \citep{lavvas17} and a vertical distribution based on results of 1-D models of photochemical aerosol formation using CARMA \cite{toon1979,ackerman1995}.  We additionally ran several simulations with more reflective hazes (specifically, tholins \citep{khare84} and an idealized conservative scattering haze) due to our finding that GJ\,1214b has a high Bond albedo.  From these simulations, we forward-modeled synthetic MIRI phase curves and phase-resolved spectra for comparison to the data (see Figures~\ref{fig:wl_models} \& \ref{fig:fp_day_night}).

We found that the broadband MIRI phase curve is best matched by GCM simulations that include a metallicity in excess of 100$\times$ solar and a thick haze composed of highly reflective aerosols (Figure~\ref{fig:wl_models}).  The high metallicity is required to produce the large observed phase curve amplitude \cite[][Extended Data Figure~\ref{fig:amp_offset}]{kataria14,charnay15b,charnay15,christie22}.  On the dayside, the observed phase curve falls between models that assumed tholin-like aerosols and those with purely conservative scatterers, implying single scattering albedos (SSA) in excess of $\sim$0.8 over the wavelength range of GJ\,1214's peak luminosity. 
 Clear atmospheres absorb too much stellar radiation and are thus all globally too hot to match the observed phase curve.  Furthermore, the nominal CARMA haze model did not provide sufficient radiative feedback to significantly alter GJ\,1214b's thermal structure; models that best match the data required a haze optical thickness enhanced by a factor of 10.  

A consistent picture of a planet with a high metallicity atmosphere and a thick, high-albedo haze qualitatively agrees with all of the available data products: the white light phase curve (Figure~\ref{fig:wl_models}), the dayside and nightside spectra (Figure~\ref{fig:fp_day_night}, discussed further, below), the phase curve amplitudes and peak offsets (Extended Data Figure~\ref{fig:amp_offset}), and the transmission spectrum (Extended Data Figure~\ref{fig:transmission}).  The latter reveals a flat spectrum across the MIRI bandpass.  Due to the computational cost of running GCMs we were unable to fully sample the possible parameter space, and we expect that a finer sampling and perhaps introducing non-uniform aerosol coverage would serve to further refine our best-fit results.

%Our best-fit 3-D forward models when compared against the MIRI phase curve data have metallicity in excess of 100$\times$ solar, haze that is approximately 10$\times$ more abundant than our nominal model, and haze particles with a high SSA. The high haze SSA implies more reflective aerosols on GJ\,1214b than what are found in Titan's atmosphere. This picture of GJ\,1214b's atmosphere qualitatively agrees with all of the available data products: the white light phase curve (Figure~\ref{fig:wl_models}), the dayside and nightside spectra (Figure~\ref{fig:fp_day_night}, discussed further, below), the phase curve amplitudes and peak offsets (Extended Data Figure~\ref{fig:amp_offset}), and the transmission spectrum (Extended Data Figure NN).  The latter reveals a flat spectrum across the MIRI bandpass.  Due to the computational cost of running GCMs we were unable to fully sample the possible parameter space, and we expect that a finer sampling and perhaps introducing non-uniform aerosol coverage would serve to further refine our best-fit results.

% 

Finally, we investigated whether the shape of GJ\,1214b's dayside and nightside spectra can reveal anything about the planet's atmospheric composition.  The planet's dayside spectrum (Figure~\ref{fig:fp_day_night}, panel a) appears blackbody-like by eye, with a best-fit temperature of $553 \pm 9$\,K.  However, a more careful analysis reveals that the dayside spectrum is formally discrepant from an isothermal atmosphere at the 3.1$\sigma$ confidence level (see Methods).  The nightside spectrum (Figure~\ref{fig:fp_day_night}, panel b) is even more inconsistent with a single-component blackbody (at $>6\sigma$), with a deep absorption feature clearly visible at \mbox{5\,--\,9\,$\mu$m}.  Interestingly, this feature is also mirrored in the GCM-derived spectra, which arises in the models from overlapping bands of H$_2$O (5\,--\,8\,$\mu$m) and CH$_4$ \mbox{(7\,--\,9\,$\mu$m)}.

%determine whether any formal detections of atmospheric species are implied. 

%The nightside retrievals additionally produce weak detections of CH$_4$ (1.6$\sigma$) and HCN (1.7$\sigma$). 

We ran retrievals on the dayside and nightside spectra to obtain further constraints on the planet's atmospheric composition. When removing the data at $\lambda > 10.5$\,$\mu$m due to concerns about correlated noise at these wavelengths (see Methods), our dayside and nightside free retrievals identify H$_2$O as the absorber at the 2.5$\sigma$ and 2.6$\sigma$ confidence levels, respectively (Extended Data Figure~\ref{fig:retrievals}). There is some evidence of additional absorption on the nightside from a combination of CH$_4$ and HCN (identified at 1.6$\sigma$ and 1.7$\sigma$ confidence, respectively). Interestingly, HCN is predicted to form as a byproduct of CH$_4$ photolysis; the latter process is needed to catalyze the formation of hydrocarbon haze \cite{kempton12}. We caution that we consider all of these molecular detections to be tentative because of the low resolution and signal-to-noise of the data and having not performed an exhaustive search of all possible absorbers.  In particular, we have found that minor changes in the shape of the spectrum around the  5\,--\,9\,$\mu$m feature can result in a tradeoff between detections of H$_2$O, CH$_4$, and HCN.  One of these molecules is always retrieved at high abundance, independent of choices in the data reduction, and when taken together we find here that H$_2$O$+$CH$_4$ (H$_2$O+HCN) are jointly detected at the 3.0$\sigma$ (3.1$\sigma$) level in the nightside spectrum.  Our retrievals allow for water abundances in excess of 100$\times$ solar, which is consistent with the results from our GCM investigations.

While direct spectroscopic characterization of GJ\,1214b's atmosphere remains challenging due to thick and pervasive aerosol coverage, the JWST MIRI phase curve provides considerable insight into the planet's atmospheric properties.  Our finding that GJ\,1214b's atmosphere is highly enhanced in heavy elements relative to its host star rules out the scenario of an unaltered primary atmosphere.  Instead, our observations are consistent with a hydrogen-rich but metallicity-enhanced atmosphere. This is in line with predictions that sub-Neptunes retain primordial atmospheres composed of nebular gases, which are sculpted by mass loss that is either photoevaporative or core-powered \cite{owen13,gupta19}.  Alternatively, given the possibility of high water abundance from our retrievals and our inability to rule out very high metallicity atmospheres, our results are also consistent with a  ``water world" planetary scenario, in which the planet's high \textit{bulk} water content would result from formation beyond the water ice line or incorporation of significant material from that region  \cite{kuchner03,leger04,rogers10b}.  

The high observed Bond albedo of GJ\,1214b opens new questions as to the nature of the planet's aerosol layer.  The previous candidates for hydrocarbon hazes (i.e., soots and tholins) are too absorptive to match our observations.  Laboratory experiments that generate photochemical hazes for sub-Neptune-like environments hint at a wider variety of outcomes for haze formation \cite{horst18,he18,gavilan18} and could provide viable candidates.  Reflective clouds such as KCl and ZnS are also a possibility, but it has been challenging to find scenarios that form sufficiently thick clouds high enough in the atmosphere to match the featureless nature of GJ\,1214b's transmission spectrum \cite{morley15, ohno18}.  The high observed albedo of GJ\,1214b motivates further work toward understanding the diversity of aerosols that can exist in exoplanetary environments.

\section*{Methods}

\renewcommand{\tablename}{Extended Data Table}
\setcounter{table}{0}

\renewcommand{\figurename}{Extended Data Fig.}
\setcounter{figure}{0}

\subsection*{Data reduction}
Our primary data reduction was carried out using a new, end-to-end pipeline (\texttt{SPARTA}) that begins with the raw, uncalibrated files. This pipeline was tested on the L\,168-9b transit observation obtained during JWST commissioning and we found agreement with the results of Ref.~\cite{bouwman22}. We also performed independent reductions of the GJ\,1214b phase curve using the \texttt{Eureka!} package \cite{eureka} as a further check of the robustness of our results. The \texttt{Eureka!} package uses the stage 1 and 2 reduction routines from the JWST Science Calibration Pipeline. We found that we could get consistent results between \texttt{SPARTA} and \texttt{Eureka!} when adopting the same assumptions for the exponential ramp used to model the instrument systematics (see below). We ultimately chose the reduction described in detail here as the primary result for this paper because it gave the smallest scatter of the residuals (significantly outperforming the other reductions at wavelengths $<8\,\mu$m), it was more robust to variations in the reduction and analysis parameters, and it was more extensively developed and tested for this data set.

The raw data files for the GJ\,1214b observation contain 42 up-the-ramp samples (``groups'') for every integration, row (the spatial direction), and column (the spectral direction).  They are therefore 4D arrays.  The phase curve was broken up into five exposures due to data volume limitations. The breaks between the exposures were 40\,s, during which the telescope maintained fine guidance pointing. STScI further divided the delivered data for the exposures into ten segments each to keep file sizes reasonable, resulting in 50 uncalibrated files.

First, we redivide segments 5 and 6 of exposure 3 into three segments: one with only pre-transit data, one that covers the transit only, and one with only post-transit data.  This way, each segment has similar flux throughout, with the exception of the short ingresses and egresses in the middle segment.

Second, we calibrate the data using the reference files provided by STScI: version 32 of the nonlinearity coefficients, version 84 of the dark, version 789 of the flat, version 85 of the read noise, version 73 of the reset, and version 6 of specwcs.  We subtract the reset anomaly, apply the non-linearity correction, subtract the dark current, and multiply by the gain, in the same way as the official JWST pipeline. 

The gain value currently provided in JDocs and the calibration reference files (5.5 e-/DN) is known to be incorrect. The MIRI detector gain was found to be wavelength-dependent, varying from $\sim$2.9\,e-/DN at 5\,$\mu$m to $\sim$3.6\,e-/DN at 15\,$\mu$m and beyond (STScI, private communication). We adopt here an intermediate value of 3.1\,e-/DN, which is consistent with the values adopted in the JWST ETC over the MIRI LRS passband. 

In addition, the non-linearity correction was found to leave higher than desired residuals, especially in the brightest pixels (of order 100s of DN). Currently, the official JWST reference file reports the same correction parameters for all pixels, whereas further investigation suggests that the pixels' non-linearity behavior displays a flux dependency. This issue is being investigated further at STScI.  Therefore, we adopt a two-step process to fit the up-the-ramp samples and obtain slopes for each pixel of each integration.  For each file (corresponding to one segment of one exposure), we:

\begin{enumerate}
    \item Fit the up-the-ramp samples for each pixel of each integration.
    \item Calculate residuals of these fits.
    \item Calculate the median residuals for every group and pixel, across all integrations in the file.
    \item Subtract these median residuals from the original data
    \item Fit the up-the-ramp samples for each pixel of each integration again.  Groups with values that deviate from the fit by more than 5$\sigma$ are rejected, and the fit is repeated.  Approximately 0.032\% of all groups are flagged as bad.  This is repeated until convergence.
\end{enumerate}

Fitting a line to up-the-ramp data is not as trivial as it might appear, because each sample has two sources of noise: read noise, which can be assumed to be independent, and photon noise, which depends on all of the photons accumulated so far and is thus not independent.  The problem is simpler when the differences between adjacent reads are considered.  For the differences, there are two sources of noise: photon noise for the photons accumulated between the two reads, which is independent, and read noise, which is correlated with the read noise on the previous difference.  Due to this covariance, there is no simple method of optimally estimating the slope.  The other naive method of subtracting the first read from the last and dividing by the interval is optimal in the limit where photon noise far exceeds read noise, while the other method of fitting a line to all reads with equal weights is optimal in the limit where read noise far exceeds photon noise.  The optimal estimate can only be obtained by considering the covariance matrix of the differences \citep{fixsen_2000}:

\begin{align}
\begin{bmatrix}
(s + 2R^2) & -R^2 & 0 & 0 & ...\\
-R^2 & (s + 2R^2) & -R^2 & 0 & ...\\ 
0 & -R^2 & (s + 2R^2) & -R^2 & ...\\
...
\end{bmatrix}
\end{align}
where $s$ is the signal in photons/group, and $R$ is the read noise.  The inverse of this matrix is the precision matrix, and the sum of the precision matrix's columns (or rows; the matrix is symmetric) gives the optimal weights to apply to the first differences so that their weighted average optimally estimates the slope.  The variance on the estimate is then the inverse of the sum of the weights.

To invert the covariance matrix, we note that it is a tridiagonal matrix with a constant diagonal and constant off-diagonals.  This type of matrix has an analytical inverse, given in Equation 10 of \cite{hu_1996}.  We sum over the rows of the inverse with the help of Mathematica and obtain
\begin{equation}
    l \equiv \mathrm{arccosh}(1 + \frac{s}{2R^2})\end{equation}
\begin{equation}
    w_j = -\frac{e^l (1-e^{-jl}) (e^{(jl-lN)} - e^l)}{R^2 (e^l - 1)^2 (e^l + e^{-lN})}\\
\end{equation}
for the weight of the $j$th first difference out of $N$.  The sum of $w_j$ over $j$ is also analytic, but its expression is longer, so we numerically compute the sums.

%\begin{align}
%    l &\equiv arccosh(1 + \frac{s}{2R^2})
%    w_j &= -\frac{e^l (1-e^{-jl}) (e^{(jl-lN)} - e^l)}%{R^2 (e^l - 1)^2 (e^l + e^{-lN})}\\
%\end{align}

After generating the rateints files, we measure the position of the trace in each integration.  We compute a 2D template by taking the pixel-wise median of all integrations.  To get the position of the trace in any given integration, we shift and scale the template until it matches the data, using \texttt{scipy's} Nelder-Mead minimizer to find the optimal parameters.  The template matching algorithm ignores the 10 rows closest to either edge, and only consider columns 26--46 (that is, the 10 closest to the trace).  We ignore trace rotation.  Although the inferred ``scale'' (the number which multiplies the template) could be used directly as the flux, we decided to have a dedicated spectral extraction step and consider the scale as a nuisance parameter.

After measuring trace positions, we perform spectral extraction.  For each row between 141 and 386 inclusive, corresponding to 5\,--\,12\,$\mu$m, we sum the columns 33\,--\,39 inclusive.  This 7-pixel extraction window is centered on the brightest column, namely 36.  The background is computed by averaging columns 10:24 and 48:62 (inclusive and zero-indexed) of each row of each integration (i.e., two 15-wide windows equidistant from the trace, one on each side of the trace) and subtracting the result from the flux on a per-integration, per-row basis.  We do not repair any bad pixels at this stage because our attempts have resulted in minimal or deleterious effects.

As the final step of the reduction, we gather all 21,600 fluxes and positions, and identify the bad integrations.  The integrations where the minimizer failed to find a position offset are flagged as bad.  We compute a detrended version of the light curve by subtracting a median-filtered version with a kernel of size 216.  Integrations that are more than 4$\sigma$ from zero in this detrended light curve are flagged as bad.  The first 10 integrations in the observation are also always marked as bad.  Excepting the 25 bad integrations at the very beginning, which we trim off, we find 44 bad integrations (0.2\% of the total).

\subsection*{Time-series systematics}
\begin{figure}[h]
    \centering
\includegraphics[width=\linewidth]{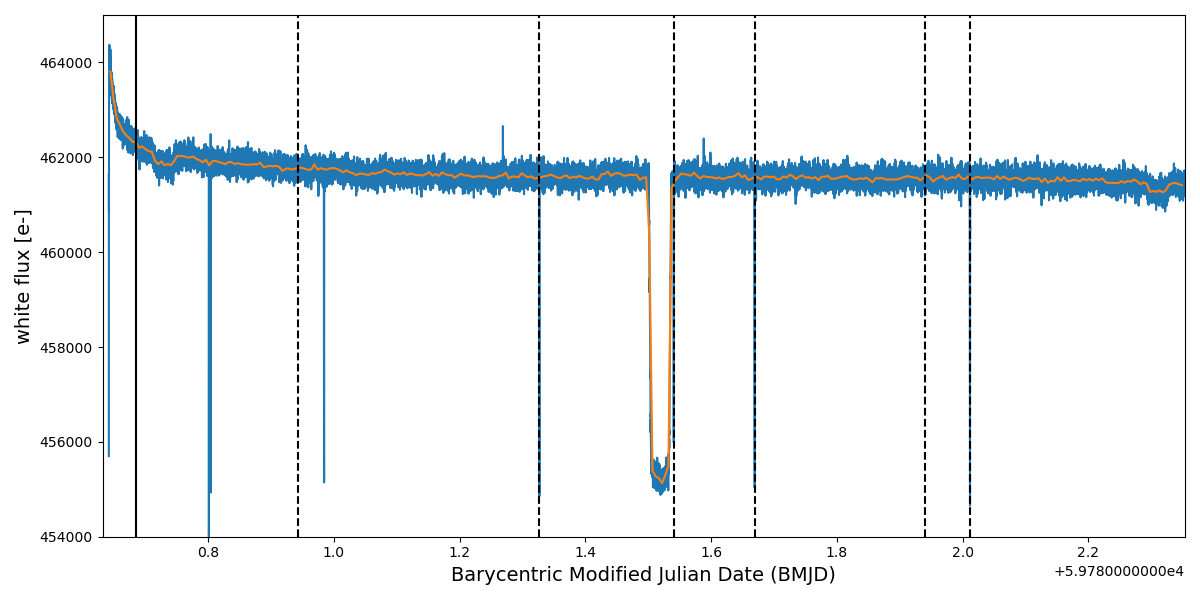}
\caption{\textbf{Raw white light curve for GJ\,1214b.} All the individual integrations are shown in blue. A median filtered (64 points) version of the light curve is shown in orange. For our analysis we discard the 550 integrations (63 min) before the vertical black line.  Note the higher discrepant integrations, some of which correspond to HGA moves (vertical dashed lines); the ramp at the start of observations; and the pre-transit brightening.}
    \label{fig:raw_white_lc} 
\end{figure}

We inspected the white light curve (Extended Data Figure~\ref{fig:raw_white_lc}) resulting from the reduction described above to identify instrument systematics in the data. We find that the time-series exhibits a ramp downward with a variety of e-folding timescales (roughly 5\,--\,90 minutes), of which the shortest timescales are most visible early in the visit.  The white light curve also exhibits a linear drift in time of $\sim$1300\,ppm from one secondary eclipse to the next, partially explained by a linear drift in the $y$-position of the spectrum (i.e., the dispersion direction) of $\sim$0.033 pixels.  
Additional correlated (red) noise is apparent in both the white light and spectroscopic light curves.  It is not clear what the exact noise sources are, but we suspect that they are primarily instrumental. We refer the reader to Morrison et al.\ (submitted), where the various detector systematics that are likely to be present in our data are described in more detail.   We also find that the time-series exhibits a mysterious 200\,ppm pre-transit brightening starting at phase -0.06 (see orange curve in Extended Data Figure \ref{fig:raw_white_lc}), with no obvious wavelength dependence.  This brightening is seen in all independent reductions.  It is unclear whether the brightening is planetary, stellar, or instrumental.  Finally, there were six high gain antennae (HGA) moves during the observation, four of which led to a momentary decrease in flux (Extended Data Figure \ref{fig:raw_white_lc}).

The JWST observation was obtained during a time of maximum brightness for GJ\,1214 over the last five years \citep{henry23}, suggesting a period of minimal spot coverage. The rotation period of the star is also estimated to be $>$50\,days, which is much longer than the timespan of the phase curve. No spot crossings are seen in the transit. Therefore, we do not expect stellar activity to have impacted the data.

\subsection*{Fitting the time series}
We infer system parameters from the light curves with \texttt{emcee} \cite{foremanmackey13}.  The free parameters in the white light curve fit are the transit and eclipse time, the transit and eclipse depth, the scaled semi-major axis ($a/R_{s}$), the impact parameter ($b$), the nominal stellar flux ($F_\star$), a coefficient that multiplies the $y$-position of the trace ($c_y$), the amplitude ($A$) and timescale ($\tau$) of the exponential ramp, an error factor that multiplies the nominal errors, and the four sinusoidal coefficients of a second order phase curve ($C_1$, $D_1$, $C_2$, $D_2$). 

\begin{figure}[h]
    \centering
    \includegraphics[width=0.9\textwidth]{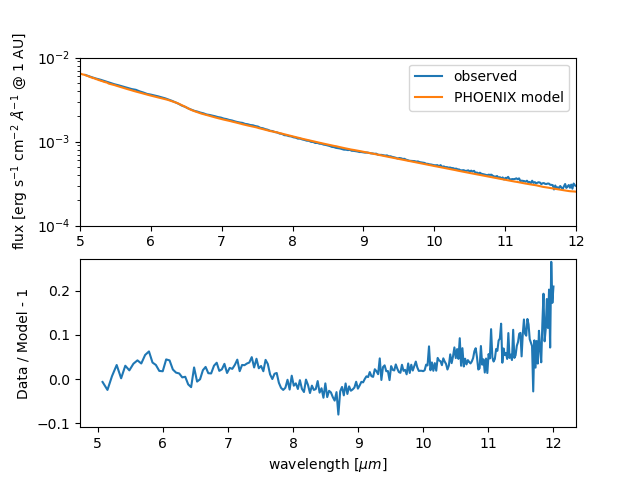}
    \caption{\textbf{Observed stellar spectrum, compared to the PHOENIX model we adopted.}  The top panel shows the modeled and observed spectra. 
 The bottom panel shows the residuals as a ratio.}
    \label{fig:stellar_spec_comp} 
\end{figure}

To avoid having to fit the steep ramp at the beginning of the observations, we discard the first hour (550 integrations) of observations.  We also identify points that lie more than 4 sigma from the median-filtered light curve (filter size: 216 points, or 1\% of the total length) as outliers and reject them.  Depending on the wavelength, we reject between 0 and 17 points (0--0.08\% of the total). The transit is modelled with \texttt{batman} \cite{batman}, assuming a circular orbit with a period of 1.58040433\,d \cite{cloutier21}, and limb darkening coefficients computed from a \texttt{PHOENIX} model.  The \texttt{PHOENIX} model has parameters $T_{\mathrm{eff}}$ = 3250\,K, $log\,g = 5.0$, and [M/H] = +0.2 and is part of the set of the models originally used in Ref.~\cite{kreidberg14}. This same stellar model is used throughout the rest of our modeling and interpretation of the data, and for error propagation we adopt an uncertainty on $T_{\mathrm{eff}}$ of $\pm$100\,K \cite{cloutier21}.  The disk-integrated spectrum computed in this model provides a good match to the flux calibrated spectrum of GJ\,1214 extracted from the MIRI data, as shown in Extended Data Figure \ref{fig:stellar_spec_comp}.

The systematics model is:
\begin{align}
    S = F_* (1 + A\exp{(-t/\tau)} + c_y y + m(t-\overline{t}))
\end{align}
The planetary flux model is:
\begin{align}
    F_p = E + C_1(\cos(\omega t) - 1) + D_1 \sin(\omega t) + C_2(\cos(2\omega t) - 1) + D_2 \sin(2\omega t)
    \label{eq_flux}
\end{align}
where E is the eclipse depth, $t$ is the time since secondary eclipse, $\overline{t}$ is the mean time, and $\omega=2\pi/P$.  The -1 is included so that at the time of secondary eclipse, the planetary flux is E.  The derived planetary flux model parameters for the white light and spectroscopic phase curves are given in Extended Data Table~\ref{tab:phase_curve_params}.

The $y$-position (i.e., dispersion direction) changes nearly linearly with time during the observations, but with significant high-frequency fluctuations.  To reduce degeneracy, we subtract out the linear trend from $y$ so that all of the linear dependence of flux on time goes into the $m(t-\overline{t})$ term.  The $y$ term is important: without it,  the scatter increases by tens of percent for both the white light curve and the spectral light curves.  The $x$-position (i.e., spatial direction) doesn't carry a similar sensitivity, because small shifts in the spatial position of the trace have nearly imperceptible impacts on the measured flux.

The transit parameters found by our fit to the white light curve are shown in Extended Data Table \ref{tab:transit_params}.  The transit time is within 3.7 seconds (0.4$\sigma$) of that predicted by the ephemeris of Ref.~\cite{kokori_2022}.  The $a/R_*$ is similarly within 1$\sigma$ of that reported by Ref.~\cite{cloutier21}, although our $b$ is smaller than their value of $0.325 \pm 0.025$ by 1.4$\sigma$.  The eclipse in our data happens $80\pm 16$ seconds after phase=0.5.  Fifteen seconds of the delay can be explained by light travel time, leaving $65 \pm 16$ s unexplained.  If this delay is due to eccentricity and not underestimated errors resulting from systematics, it would imply $e\cos(\omega) = 0.00075 \pm 0.0004$.  This is consistent with $e\cos(\omega)=-0.007_{-0.023}^{+0.032}$ derived by Ref.~\cite{gillon14} from Spitzer eclipse observations.  Alternatively, hot spot offsets in planetary atmospheres can cause apparent timing delays, and we estimate that GJ\,1214b's observed hot spot offset could cause a delay consistent with the one we observe.  Due to our overall degree of consistency with the system parameters from Ref.~\cite{cloutier21}, we ultimately adopt their values and associated error bars for $a$, $R_p$, and $R_*$, as well as the size ratios $a/R_*$ and $R_p/R_*$ in our subsequent theoretical modeling efforts and for all calculations that rely on these parameters.  

To fit the spectral light curves, we fix the transit time, eclipse time, $a/R_{s}$, and $b$ to the best fit values found in the white light curve fit.  All other free parameters in the white light curve are also free in the spectral fit.  The limb darkening parameters are again computed from the PHOENIX model.  For wavelengths larger than 10\,$\mu$m, the noise is large enough that the timescale of the exponential ramp is poorly constrained, and large timescales become degenerate with the phase curve parameters because, combined with $c_y y$, they can mimic the phase curve.  We therefore give the timescale a Gaussian prior with mean 0.035\,d and standard deviation 0.01\,d for the three reddest wavelength bins, spanning 10.5\,--\,12.0\,$\mu$m.  The prior of 0.035\,d was chosen because it is similar to the timescales at 9.0\,--\,10.5\,$\mu$m.  

The white light fit achieves a RMS of 280\,ppm, which is 12\% above the photon noise if the gain is assumed to be 3.1 electrons/data number.  In the spectroscopic channels, the RMS of our residuals is 6\% above photon noise in our bluest bin (5.0\,--\,5.5\,$\mu$m), dropping to 0.5\% above photon noise at 8\,$\mu$m before rising again to 13\% in the reddest bin (11.5\,--\,12.0\,$\mu$m).  We note again that the gain is wavelength dependent and has not yet been finalized, so these values are only accurate to several percent.

We performed many tests to assess the robustness of the results from our primary reduction, most of which produced negligible changes ($\lesssim 0.5\sigma$ for every wavelength in the transit spectrum, emission spectrum, nightside emission spectrum, phase curve amplitude, and phase curve offset).  For example, we tried MCMC chain lengths of 3,000 and 30,000, and obtained identical results.  We tried ignoring the pre-transit anomaly by masking phases -0.06 to -0.011, finding minimal differences even in the nightside emission spectrum.  We tried decorrelating against the trace's $x$-position, which changed nothing because the vast majority of the flux was already within our window, and the pointing was very stable.  However, some of our tests resulted in small changes (generally $\sim$0.8$\sigma$ shifts at a few wavelengths).  These include using optimal extraction, shrinking the aperture half-width from 3 to 2 pixels, and not ignoring the first five groups for the first round of up-the-ramp fitting.  The last test resulted in transit spectra $\sim$100\,ppm lower, probably due to the reset switch charge decay \citep{argyriou_phd}.

One test that resulted in more substantial changes was using different regions to estimate the background.  Using the rightmost 15 pixels instead of two 15-pixel windows on either side of the trace results in a declining nightside emission spectrum beyond 10\,$\mu$m instead of a flat spectrum, opening up a gap of 300\,ppm at the reddest wavelengths.  The former approach results in fewer
artifacts in the spectral light curves, but the latter approach results in far cleaner background-subtracted 2D images, making it unclear which approach is best.  Ultimately, we chose the latter approach.  This finding that the nightside spectrum at the reddest wavelengths depends on choices in background subtraction led us to perform our retrievals only on the data shortward of 10.5\,$\mu$m, as detailed in the main text. 

Among all our tests, the ones that most significantly affect our results are those relating to the ramp.  The ramp at the beginning of the observations is degenerate with the phase curve.  Intuitively, any curvature in the light curve can be attributed to either the ramp or the phase curve.  Since we have no independent model of the ramp and do not know its exact functional form, it is difficult to know what to attribute to the ramp and what to attribute to the phase curve. 

When we trim only 30 minutes instead of 60 minutes from the beginning of the observations, both the white light fit and the bluer ($<$8\,$\mu$m) spectroscopic fits strongly prefer negative planetary fluxes for the coldest hemisphere.  The inferred exponential decay timescale is shorter, likely because the ramp has components with many different timescales and the short-timescale contributions are suppressed with more aggressive trimming.  If we fix the timescale to the value found in our fiducial fit, we fail to fit the very rapid decline in flux at the very beginning.  If we fit two exponential ramps instead of one, we obtain larger error bars on inferred parameters, but do not resolve the problem of the fit preferring negative fluxes.  We can resolve the problem by imposing Gaussian priors on the amplitude and timescale of the ramp, but we had no physical justification for these priors. 

In the end, we decided to trim as much data as we could before the first eclipse in order to eliminate as much of the ramp as possible, and assume that the remainder is accurately modelled by a single exponential.  Across all of our tests, the dayside spectrum and the shape of the nightside spectrum shortward of 10\,$\mu$m remain consistent.  The choice of ramp parameters impacts the phase curve primarily by altering the (absolute) nightside flux and therefore the phase curve amplitude, and also the phase curve offset.  Across all of our reductions using reasonable choices for trimming, fitting the ramp, and background subtraction, we find that the phase curve amplitude, offset, and nightside planet-star flux ratio differ by up to 17 ppm, 7$^\circ$, and 40 ppm, respectively in the white light phase curves.  The large uncertainty on the nightside flux, in particular, impacts our estimates of GJ\,1214b's Bond albedo; derived albedos from our various data reductions give values between 0.39 and 0.61.  This implies that our formal error bar on the Bond albedo reported the main text may be underestimated by a factor of up to $\sim$2.  Additionally, as with all phase curve observations, the peak offsets (Extended Data Figure~\ref{fig:amp_offset}) are quite sensitive to the treatment of time-varying systematics and therefore may also have larger uncertainty than the formal error bars suggest.  Ultimately a better understanding of the origin and nature of the ramp is necessary for improving confidence in these derived phase curve parameters.

\subsection*{Temperature Map}

We followed \cite{cowan2008} in producing a longitudinal brightness map from the white light curve. The planetary flux given by Equation~\ref{eq_flux} is converted to longitude-based sinusoids 
\begin{align}
    \frac{F_p(\phi)}{F_*} = A_0 +A_1cos(\phi)+B_1sin(\phi)+A_2cos(2\phi)+B_2sin(2\phi)
\end{align}
where $\phi$ is longitude, and 
\begin{align}
\begin{array}{l}
A_0 = (F_p-C_1-C_2)/2 \\
A_1 = 2C_1/\pi \\
B_1 = -2D_1/\pi \\
A_2 = 3C_2/2 \\
B_2=-3D_2/2.
\end{array}
\end{align}
We then inverted $F_p(\phi)$, assuming blackbody emission, to obtain the corresponding longitudinal brightness temperature curve (Figure~\ref{fig:T_map}, panel a).  By assigning a $\cos(\theta)$ weighting of the planetary flux with latitude, $\theta$, we plot the brightness temperature map (Figure~\ref{fig:T_map}, panel b) using a Robinson projection.  The black region on the map is where the planetary flux is negative. 

%We then estimated $F_p/F_s$ assuming $F_p$ equals blackbody radiation with temperature ranging from 0 to 1000 K and $F_s$ from PHOENIX model. $F(\phi)/F_s$ was interpolated to $Fp_{black body}/Fs$ to calculate the corresponding longitudinal brightness temperature curve (Fig.?? Left). By assigning a $\cos(latitude)$ weight, we plot the brightness temperature map (Fig.?? Right) using Robinson projection.  The black region on the map is where the planetary flux is negative. 

\subsection*{Bond Albedo Calculation}

To compute the Bond albedo of GJ\,1214b, we need to answer four questions:

\begin{enumerate}
\item How much energy per second does the planet receive from its host star?  For illustrative purposes, we calculate the luminosity the planet receives from the star.  This quantity cancels out in the end, since we directly measure $F_p/F_*$, so the error on the quantity is irrelevant. We adopt $2.48 \times 10^{26}$ erg/s.
\item How much luminosity does the planet radiate from 5 to 12\,$\mu$m?  We measure $F_p/F_*$ as a function of phase and wavelength.  We can derive $F_p$ as a function of phase and wavelength because the stellar spectrum is known fairly accurately.  At each phase, we integrate across 5\,--\,12\,$\mu$m to calculate what the planet luminosity would be if it were isothermal.  Since the planet is not actually isothermal, we then take the mean across all phases.  Our result is $7.1 \pm 0.6 \times 10^{25}$ erg/s, where the error bar is derived from the MCMC samples.
\item What fraction of total planet luminosity is within 5\,--\,12\,$\mu$m?  For a blackbody, this fraction is close to 50\% for a wide range of temperatures: it is 48\% at 350 K, 57\% at 500 K, 54\% at 600 K, and 49\% at 700 K.  We also computed this value for our GCMs to estimate the impact of the non-isothermal nature of the energy output, and find that for almost every GCM, the value is 50--60\%.  We adopt $54 \pm 4$\% as our fiducial value.  With this assumption, the total planet luminosity becomes $1.31 \pm 0.15 \times 10^{26}$ erg/s.
\item What is the ratio of the planet flux as seen at infinite distance averaged over the equatorial plane (which is effectively our own viewing geometry), and the planet flux as seen at infinite distance averaged over all angles (which is what is needed to determine the true planet luminosity)?  We can approximate this value by considering the case of zero heat redistribution, in which case the flux the planet radiates would equal the flux it receives from the star at every latitude and longitude.  The emitted flux would therefore be proportional to $\cos(\theta)$.  Further assuming isotropic emission, and integrating the specific intensity, we find that the average observed flux is:

\begin{align}
F_{\rm equator} = \frac{8}{3} I_{\rm equator} (R/D)^2\\
F_{\rm avg} = \frac{\pi^2}{4} I_{\rm equator} (R/D)^2,
\end{align}
where $I_{\rm equator}$ is the specific intensity of emission from the equator.  The ratio $F_{\rm equator} / F_{\rm avg} = 32/(3\pi^2) = 1.08$ is the correction factor we are looking for.  Happily, it turns out to be a minor correction. We adopt an error of 0.01 on the correction factor.  With this correction, the total planet luminosity becomes $1.21 \pm 0.14 \times 10^{26}$ erg/s.
\end{enumerate}

The Bond albedo is then $1 - L_p/L_{in}$, where $L_p$ is from step 4 and $L_{in}$ is the incident flux from step 1.  We obtain $A_B = 0.51 \pm 0.06$.

We note that this assumes GJ\,1214b's global energy balance is dominated by the stellar irradiation and that any flux from the interior is negligible in comparison.  If the planet has a considerable intrinsic luminosity (unlikely but not possible to rule out with our data), then the Bond albedo would be even higher.

As a consistency check on the previous calculation, we also calculate the Bond albedo directly from the temperature map shown in Figure~\ref{fig:T_map}.  In this case, we calculate $L_{p}$ by directly integrating the temperature field given by the map.  This is then divided by $L_{in}$ and subtracted from unity, to give
\begin{align}
    A_B = 1 -  \frac{a^2}{\pi T_{*}^4 R_{*}^2} \iint T_{p}(\theta, \phi)^4 \sin \theta \,d \theta \,d \phi,
\end{align}
following Equations 6 and 9 from \cite{keating19}. We obtain a result of \mbox{$A_B = 0.49 \pm 0.05$}, which is consistent with the previous calculation.  We formally adopt the previous calculation of the Bond albedo because it was more directly derived from the observational data.

\subsection*{General Circulation Models}

We simulated the atmosphere of GJ\,1214b using the SPARC/MITgcm \citep{showman09,kataria13}. The model solves the primitive equations using the dynamical core of Adcroft et al. \citep{adcroft04} and is coupled to wavelength-dependent radiative transfer \citep{marley1999} using the correlated-k method in 11 wavelength bins.
In the simulations, we used the best-fit system parameters from Ref.~\cite{cloutier21}, an internal temperature of 30\,K, and the same 3,250\,K PHOENIX stellar model described in the data reduction section, above. 
%In the simulations, we used a planet radius of $1.7469282 \times 10^7$\,m, a gravity of 10.65\,m\,s$^{-2}$, an internal temperature of 30\,K, and the same 3,250\,K PHOENIX stellar model described in the data reduction section, above.
%The stellar spectrum was interpolated from a grid of PHOENIX models to an effective temperature of 3,250~K. 
%We used a ratio of the semimajor axis to the stellar radius $a/R_\star$ of 14.9, corresponding to a zero-albedo equilibrium temperature of the planet of 595~K. 
All simulations assume equilibrium chemistry abundances for the gas. For pressures greater than 10\,bar, we employ a bottom drag that linearly increases with pressure \citep{liu13}, with a maximum drag timescale of $10^{5}$\,s at the bottom layer, and we apply a Shapiro filter throughout the simulation to suppress small-scale numerical noise. Our simulations have a horizontal resolution of C32 (equivalent to 128$\times$64) and vertically extend from 200~bars to $2\times 10^{-7}$~bar, using 60 vertical layers. We used a dynamical timestep of 25~s/10~s and a radiative timestep of 50~s/20~s for metallicities up to 100$\times$~solar/300$\times$~solar and above, respectively. The initial temperature profiles were calculated with the 1D radiative transfer code HELIOS \cite{malik17,malik19}.  All GCMs were run for 1,000~simulation days. 

It has previously been shown that mean molecular weight has a leading order effect on day-night heat transport in tidally locked exoplanet atmospheres, with low mean molecular weight atmospheres (e.g., solar composition) having the most efficient heat transport and therefore producing the smallest phase curve amplitudes and largest peak offsets \cite{zhang17}. Prior 3-D modeling of GJ\,1214b has affirmed this trend in the sub-Neptune regime \cite{kataria14} and has furthermore shown that condensate clouds only moderately perturb the clear atmosphere expectations \cite{charnay15b,charnay15,christie22}.  Thick photochemically-derived hazes however, such as are expected to be present in GJ\,1214b's atmosphere based on prior transmission spectroscopy observations, have not been modeled in GCMs previously.  We include such haze layers in our modeling here, in order to understand their impact on the JWST MIRI phase curve. 

%haze parametrization
For our simulations with photochemical hazes, we added horizontally uniform haze extinction to the model, with vertical profiles of the optical depth, single-scattering albedo and asymmetry parameter derived from the 1D microphysics model CARMA (Community Aerosol and Radiation Model for Atmospheres) \cite{toon1979,ackerman1995}. In particular, we follow the same haze modeling strategy as \cite{adams19}. Briefly, 10 nm radii spherical seed haze particles are added to the model atmosphere from the topmost model layer with a user-chosen column-integrated production rate and allowed to coagulate with each other to grow to larger sizes. Primary (monomer) haze particle sizes range between a few to a few tens of nm in the atmospheres of hazy solar system worlds \cite{tomasko2009,lavvas2010,gladstone2016}, motivating our choice of 10 nm for the radii of our initial seed particles. These particles are also transported around the atmospheric column via sedimentation and eddy diffusion, with an eddy diffusion coefficient of 10$^7$ cm$^2$ s$^{-1}$ that is constant with altitude. We base this value on the GCM simulations of Ref.~\cite{charnay15b}, The microphysics model assumed a column-integrated haze production rate of $10^{-12}$ g cm$^{-2}$ s$^{-1}$ and a background atmosphere with 100$\times$~solar metallicity. The haze production rate was chosen as a typical value derived from photochemical models \cite[e.g.,][]{kawashima19b}, though its value can vary by orders of magnitude. 
To simulate higher (or lower) haze production rates in the GCMs, we multiplied the optical depth in each layer of the atmosphere by a fixed scaling factor.  We explored three different cases for the haze optical properties: soot \cite{lavvas17}, tholins \cite{khare84}, and highly reflective hazes.  The latter were constructed to have identical properties as the soots, except that the single-scattering albedo was raised to 0.9999. 

We post-processed the GCM outputs to produce thermal emission spectra using the same plane-parallel radiative transfer code as in the GCM but with 196 wavelength bins. Details on the post-processing procedure can be found in Ref.~\citep{parmentier16}. We additionally post-processed the GCMs with a three-dimensional ray-striking radiative transfer code \cite{kempton2012constraining, savel2023diagnosing} to generate model transmission spectra (Extended Data Figure~\ref{fig:transmission}). We adapted this code to accept the same haze abundance and opacity profiles used in the GCM, employing a similar aerosol implementation to an emission spectroscopy version of this code \citep{harada2021signatures}.  All of our post-processing calculations use the planet-to-star radius ratio ($R_p/R_\star$) from Ref.~\cite{cloutier21}.

The full set of GCMs that we ran for this work are listed in Extended Data Table~\ref{tab:gcm_overview}. The 3-D thermal structures and atmospheric dynamics of these GCMs will be described in detail in Steinrueck et al.\ (in prep.).  As described above, our haze model was derived from a 1-D calculation and is therefore homogeneous around the entire planet.  Future work should entail the inclusion of spatially inhomogeneous hazes, including their radiative feedback, and transport, as well as the chemistry that leads to the formation and destruction of the haze particles.

\subsection*{Retrievals}
We performed atmospheric retrievals on the dayside and nightside emission spectra of GJ\,1214b using the \texttt{HyDRo} \citep{piette22} and \texttt{CHIMERA}  \citep{Line2013} retrieval frameworks. As described in the main text and in the Methods (``Fitting the time series''), we exclude data points at wavelengths $>$10.5~$\mu$m from the retrieval due to concerns about correlated noise in this region that arise from uncertainty in how to best describe the ramp parameters and choices in background subtraction.  

\begin{figure}[h]
    \centering
\includegraphics[width=\linewidth]{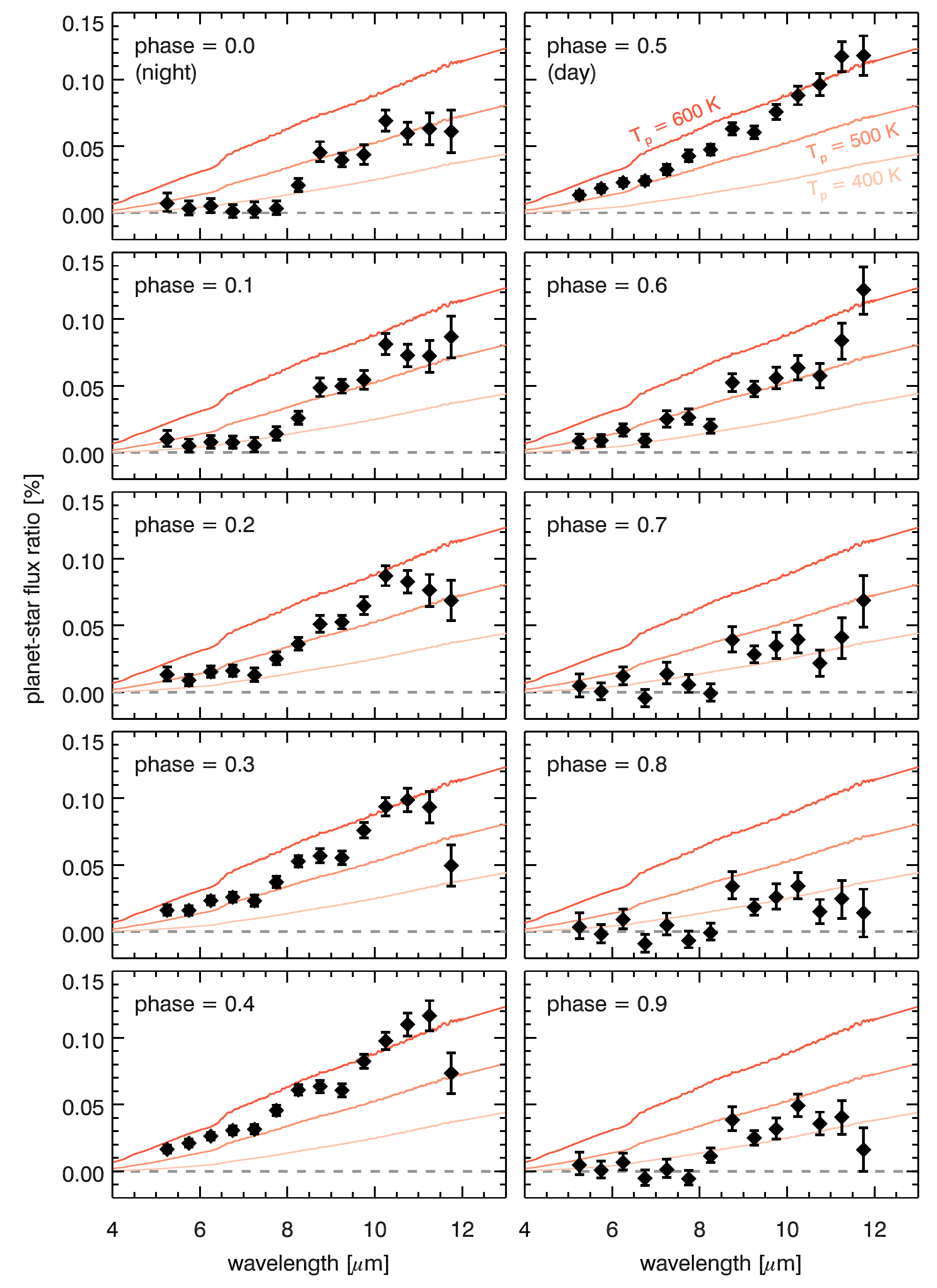}
\caption{\textbf{The observed emission spectrum of GJ\,1214b at various orbital phases.}  The upper left and upper right-hand panels correspond to the nightside and dayside emission spectrum, respectively.  Colored lines denote blackbody planetary emission at temperatures of 400, 500, and 600\,K, as indicated in the upper right-hand panel. Black points with 1$\sigma$ error bars are the wavelength-binned phase curve data.}
    \label{fig:spectra_vs_phase} 
\end{figure}

\texttt{HyDRo}, which builds on the \texttt{HyDRA} \citep{gandhi18,gandhi20,piette20} retrieval code, consists of a parametric atmospheric forward model coupled to a Nested Sampling Bayesian parameter estimation algorithm \citep{skilling06}, \textsc{PyMultiNest}\citep{feroz09,buchner14}. For each model spectrum computed in the parameter exploration, we calculate the likelihood assuming symmetric error bars on the data (calculated by averaging the positive and negative error bars in Extended Data Figure \ref{fig:spectra_vs_phase}). The atmospheric temperature profile is modeled using the parameterization of \citep{madhusudhan09}, which includes 6 parameters and is able to capture the range of temperature structures expected for exoplanet atmospheres. We investigate cases with a range of atmospheric opacity sources, including gas phase species and clouds. The gas phase opacity sources we consider are: H$_2$O\citep{rothman10}, CH$_4$ \citep{yurchenko13,yurchenko14}, CO$_2$\citep{rothman10}, HCN\citep{harris06}, NH$_3$ \citep{yurchenko11}, CO \citep{rothman10}, N$_2$ \citep{barklem16,western18} and collision-induced absorption (CIA) due to H$_2$-H$_2$ and H$_2$-He \citep{richard12}. The absorption cross sections for these species are calculated as described in \citep{gandhi17} using data from the sources cited above. We perform retrievals both with and without the assumption of a H$_2$-dominated background composition. When a H$_2$-rich background is assumed, the constant-with-depth abundance of each species other than H$_2$ or He is a parameter in the retrieval, and a solar H$_2$/He ratio is assumed. When no assumption of the background gas is made, we parameterize the abundances of each species using the centered-log-ratio (CLR) method \citep{benneke12,piette22}, which ensures identical priors for each of the chemical species in the retrieval.

The \texttt{HyDRo} retrievals also consider the effects of clouds using a simple parameterization, including the modal particle size, cloud base pressure ($P_\mathrm{b}$), pressure exponent ($\alpha$), and cloud particle abundance ($f_0$). The particle abundance is assumed to be zero below the cloud base, and to decrease at pressures below $P_\mathrm{b}$, such that at pressure $P$ the abundance is $f_0(P/P_\mathrm{b})^\alpha$. Given the temperatures probed in the atmosphere of GJ\,1214~b, KCl clouds may form on the nightside. We therefore perform retrievals with KCl clouds, using the KCl scattering and absorption properties from \citep{pinhas17}.

We also use \texttt{HyDRo} to calculate the detection significances of various chemical species. These detection significances are calculated by comparing the Bayesian evidences of retrievals which include/exclude the species in question \citep{trotta08,benneke13}. Similarly, the joint detection of two or more species can be calculated by comparing retrievals which include or exclude those species. In order to calculate the significance to which the day/nightside spectrum is inconsistent with a blackbody, we compare the Bayesian evidences of a blackbody model (with a single temperature parameter) and a simple absorption model which includes the six temperature profile parameters described above and the H$_2$O abundance (since H$_2$O is the primary absorber detected on both the dayside and nightside). We find that the observed dayside and nightside spectra are inconsistent with blackbody spectra to 3$\sigma$ and 6$\sigma$, respectively.

We perform a series of \texttt{HyDRo} retrievals on the nightside spectrum to test the sensitivity of our results to various modeling choices. We begin by testing the sensitivity of the retrieval to the species listed above, assuming a H$_2$-rich background composition. We find that the abundances of CO and N$_2$ are completely unconstrained, as expected given their minimal spectral features in this wavelength range. Furthermore, the posterior distribution for the abundance of NH$_3$ shows a strong 99\% upper limit of 10$^{-4.2}$. Given the large number of possible model parameters relative to the number of data points, we remove CO, N$_2$ and NH$_3$ from subsequent retrievals in order to minimize unnecessary parameters. While CO$_2$ was not constrained in this test, we include it out of precaution in the subsequent retrievals, as we found it to be somewhat constrained for some alternative data reductions. We also test the difference between retrievals with and without KCl clouds, finding that cloudy models are not preferred with statistical significance ($<1\sigma$ preference over the clear model). Furthermore, the posterior distributions for all other parameters are unaffected by the addition of clouds. This result does not rule out clouds (or haze) on the nightside of GJ~1214~b, but suggests that such clouds do not show significant spectral features (e.g., if the clouds are deeper than the infrared photosphere). The effects of any clouds may also be taken into account by the retrieved temperature profile which, for example, could mimic a deep cloud layer with a deep isothermal layer.

\begin{figure}
    \centering
    \includegraphics[width=0.86\textwidth]{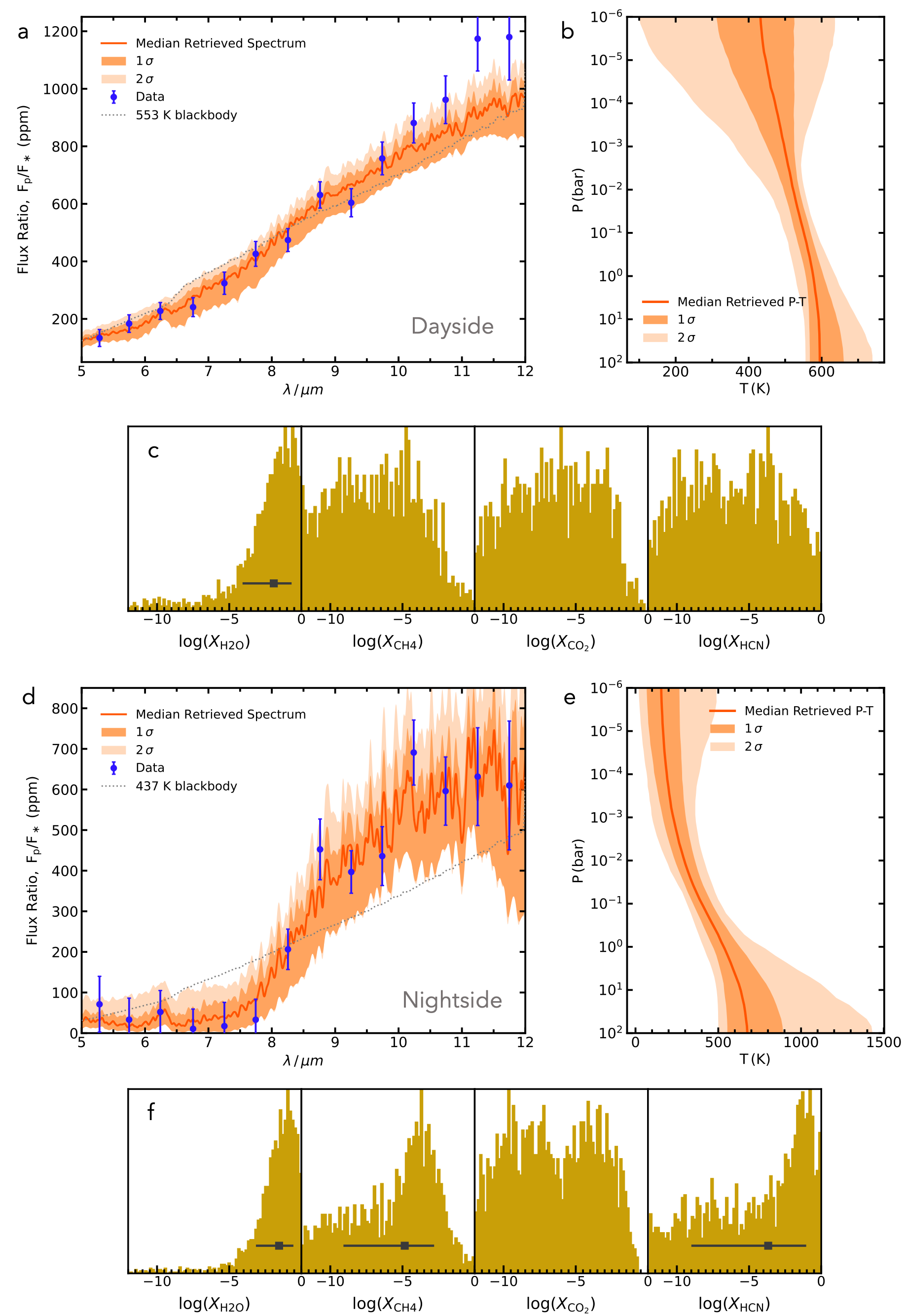}
    \caption{{\bf Dayside and nightside spectrum retrieval results obtained using the \texttt{HyDRo} atmospheric retrieval framework.} \textbf{a,d,} The best-fit retrieved spectra, and \textbf{b,e} the best-fit retrieved temperature profiles from the dayside and nightside, respectively. Dark red lines show the median retrieved spectrum and temperature profile, while dark/light shading shows the 1$\sigma$ and 2$\sigma$ contours, respectively. The blue points and 1$\sigma$ error bars in panels a and d show the observed spectra. {\bf c,f} The posterior probability distributions for the abundances of H$_2$O, CO$_2$, CH$_4$ and HCN on the dayside and nightside, respectively. The black squares and error bars show the median retrieved abundances and 1~$\sigma$ uncertainties for cases in which a bounded constraint was obtained. Only data at wavelengths $<10.5$~$\mu$m were used in the retrievals to avoid potential systematics at longer wavelengths. The retrievals are able to fit the slight absorption feature at $\lesssim 8$~$\mu$m on the dayside (panel a) with opacity from H$_2$O. The large absorption feature on the nightside at $\lesssim 8$~$\mu$m (panel d) is best fit with opacity from H$_2$O, CH$_4$ and HCN.}
    \label{fig:retrievals}
\end{figure}

We further test the effects of assuming a H$_2$-rich background compared to making no assumption about the background composition. We find that both assumptions lead to consistent results. When a H$_2$-rich background is assumed, the detection significances for H$_2$O, CH$_4$ and HCN in the nightside are 2.6$\sigma$, 1.6$\sigma$ and 1.7$\sigma$, respectively. When no assumption is made about the background composition (using the CLR method described above), the detection significances for H$_2$O, CH$_4$ and HCN are 2.5$\sigma$, 1.3$\sigma$ and 1.6$\sigma$, respectively. The tentative detection of H$_2$O is therefore robust across all retrieval models considered, while the inferences of CH$_4$ and HCN are very marginal. Extended Data Figure \ref{fig:retrievals} (panels d, e, f) shows the retrieved nightside spectrum, temperature profile and molecular abundances for our nominal \texttt{HyDRo} retrieval model, which includes H$_2$O, CH$_4$, CO$_2$ and HCN, and assumes a H$_2$-rich background.

We also perform a similar suite of retrievals on the dayside emission spectrum (Extended Data Figure \ref{fig:retrievals} panels a, b, c), and find a tentative 2.5$\sigma$ detection of H$_2$O. Similarly to the nightside, we find that NH$_3$, CO and N$_2$ are not constrained by the retrieval, and we do not find statistically significant evidence for KCl clouds (only a 1.3$\sigma$ preference for the cloudy model over the clear model). The results are consistent whether a H$_2$-rich background is assumed, or no assumption is made about the background composition.

We find that our retrieval results are broadly consistent with the inferences based on GCM models. Hazes (and clouds) are expected to affect mini-Neptune emission spectra via their radiative feedback on the atmospheric temperature profile. For example, purely scattering hazes result in more isothermal temperature profiles in 1D atmospheric models of mini-Neptunes \citep{piette20_MN}. While we do not explicitly include hazes in our retrieval models, we do include KCl clouds in our models, which have qualitatively similar effects on the spectrum as haze.  As discussed above, the clouds are neither ruled out nor statistically preferred over clear-atmosphere models, but we do retrieve a near-isothermal temperature profile for the dayside. This shallow temperature gradient may be a result of strongly reflecting hazes, in agreement with the GCM models described in the main text. Furthermore, the retrieved abundances for H$_2$O are consistent with several hundred times solar for both the dayside and nightside spectra. This is consistent with the high atmospheric metallicities inferred from the GCM models.

In order to further assess the robustness of our H$_2$O detections, we apply leave-one-out cross validation (LOO-CV) to the retrievals on the dayside and nightside spectra, following the method described in \citep{welbanks2022_loo}. We compute the expected log pointwise predictive density (elpd$_{\rm LOO}$), which quantifies the ability of the fitted model to predict unseen data, where each data point in the spectrum is left out in turn \citep{Vehtari2017}. The difference in elpd score between two models ($\Delta$elpd$_{\rm LOO}$) divided by the standard error (SE) can be used as a means of model comparison and as a complementary metric to Bayesian evidence, which is commonly used to calculate detection significances from a retrieval. Comparing models with and without H$_2$O absorption, we find that the models including H$_2$O have higher elpd$_{\rm LOO}$ scores for both the dayside and nightside spectra: $\Delta$elpd$_{\rm LOO}=2.39$ (SE =$1.46$) for the dayside spectrum and $\Delta$elpd$_{\rm LOO}=3.26$ (SE $=1.64$) for the nightside spectrum. These numbers indicate that, in both cases, the inclusion of H$_2$O absorption improves the out of sample predictive performance of the model.

We perform a second retrieval analysis with \texttt{CHIMERA} to ensure that our retrieved inferences are robust against different modelling frameworks and model prescriptions. It has been shown that, to thoroughly explore JWST observations, more than one framework needs to be used, as the precision on the observations is at the level in which model differences can be seen \citep{Barstow2020}. We performed a similar retrieval to the nominal model of HyDRo. We assume that the atmosphere is dominated by H$_2$, with a H$_2$ to He ratio of 0.17. We use the same molecules, however with a different prior assumption for each. For CHIMERA we assume a log prior from -12 to -1, hence each molecule has an upper limit of 10$\%$ of the atmosphere.  

We used a different parameterisation for the thermal structure. We use a double gray analytic temperature-pressure profile from  \cite{Parmentier2014}, which has five free parameters: $T_{\text{irr}}$, $\kappa_{\text{IR}}$, $\gamma_1$, $\gamma_2$ and $\alpha$. $T_{\text{irr}}$ is the irradiation temperature, $\kappa_{\text{IR}}$ is the infrared opacity, the parameters $\gamma_{1}$ and $\gamma_{2}$ are the ratios of the mean opacities in the two visible streams to the thermal stream: $\gamma_{1} = \kappa_{v1} / \kappa_{IR}$ and $\gamma_{2} = \kappa_{v2} / \kappa_{IR}$. The parameter $\alpha$ ranges between 0 and 1, and controls the weighting used between the two visible streams, $\kappa_{v1}$ and $\kappa_{v2}$. 

We find that our retrieved abundances and thermal structure are consistent with HyDRo within 1-sigma. This confirms that our retrieved abundances are robust against model assumptions.

\section*{Extended Data Tables and Figures}
%\newpage

\begin{figure}[h!]
    \centering
\includegraphics[width=\linewidth]{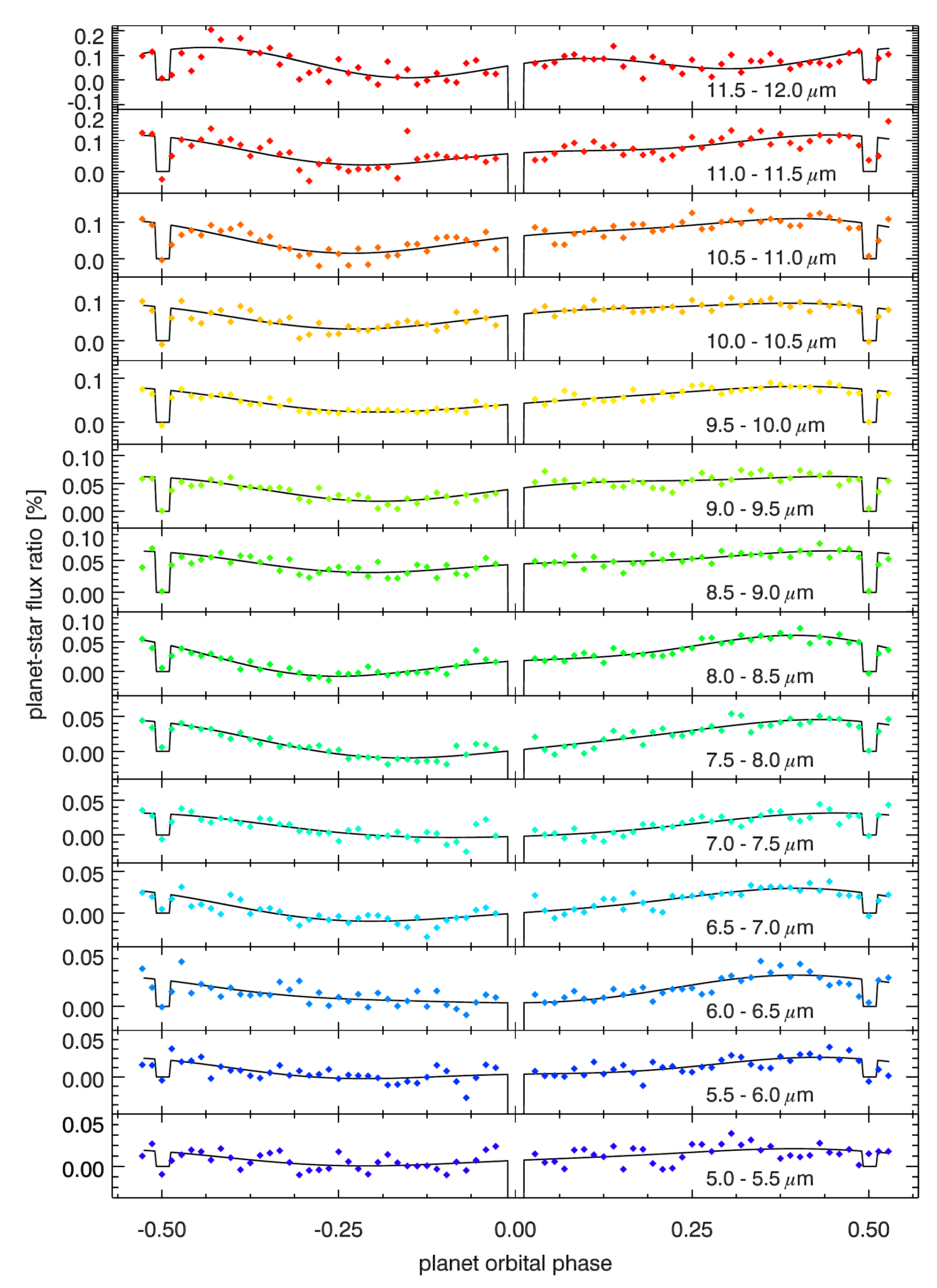}
\caption{\textbf{MIRI spectroscopic light curves from 5 to 12\,$\mu$m.} Black lines are the best-fit astrophysical model to the data, assuming a second-order sinusoid functional form for the phase variation.  Colored points are the data binned every 5 degrees in orbital phase, plotted without error bars for clarity.  Wavelength ranges for each light curve are as indicated.  Note the differing y-axis scale on each sub-panel.}
    \label{fig:spectroscopic_lc} 
\end{figure}

\begin{figure}[h]
    \centering
    \includegraphics[width=1.0\textwidth]{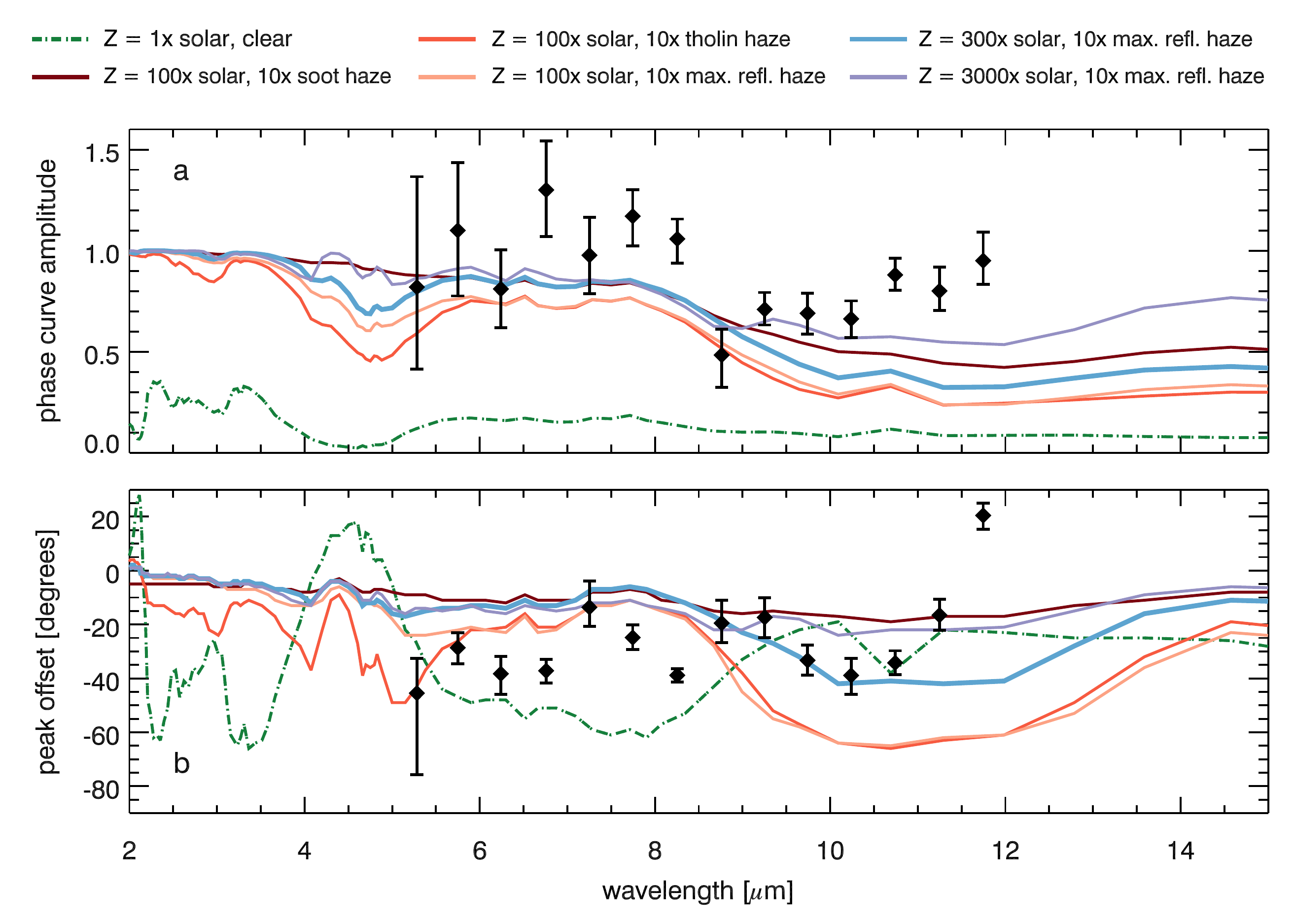}
    \caption{\textbf{Phase curve amplitudes and offsets vs.\ wavelength.} \textbf{a,} The phase curve amplitude is defined as $(F_{max} - F_{min}) / F_{max}$, where $F_{max}$ and $F_{min}$ are the maximum and minimum planet/star flux ratios from the best-fit phase curve model, respectively.  \textbf{b,} The peak offset is defined as the number of degrees in phase away from secondary eclipse at which the peak planet/star flux ratio is achieved. Negative values denote the peak occurring prior to secondary eclipse, meaning that the maximum planetary flux is eastward of the sub-stellar point.  In both panels, colored lines are the GCM-derived values for the same set of models shown in Figure~\ref{fig:fp_day_night} (see that figure's legend).  Models with higher metallicity \mbox{(i.e., $\geq$\,100$\times$ solar)} tend to provide a qualitatively better fit to the data. All error bars are 1$\sigma$.}
    \label{fig:amp_offset} 
\end{figure}

\begin{figure}[h]
    \centering
    \includegraphics[width=1.0\textwidth]{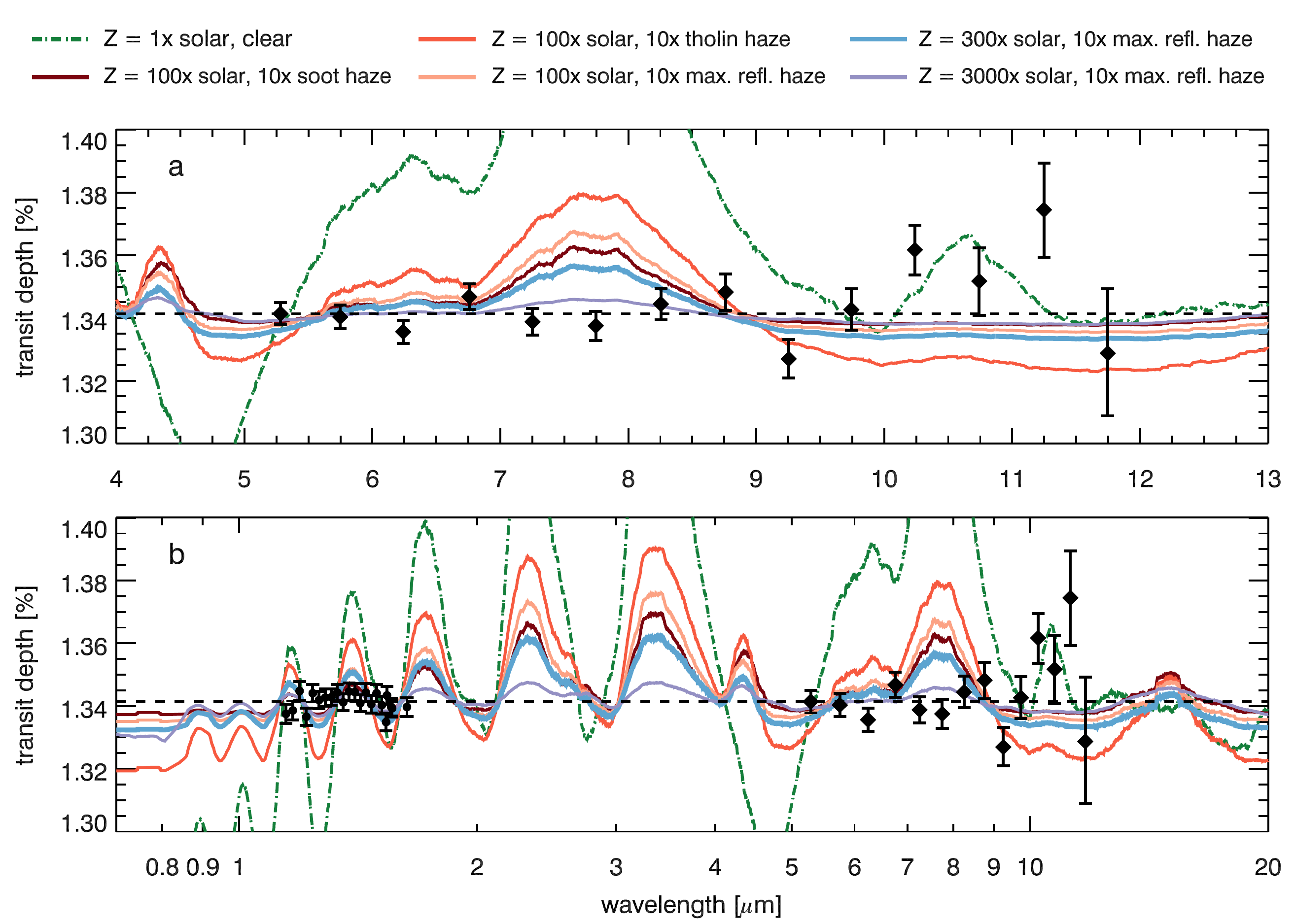}
    \caption{\textbf{The transmission spectrum of GJ\,1214b.} \textbf{a,} The MIRI data are shown compared to GCM-derived spectra from the same set of GCMs as in Figures~\ref{fig:fp_day_night} and \ref{fig:amp_offset} (see the  legend in Figure~\ref{fig:fp_day_night}). \textbf{b,} The same set of models are shown over a broader wavelength range, with the HST/WFC3 transmission spectrum from Ref.~\cite{kreidberg14} also over-plotted (smaller symbols with error bars). The WFC3 data have been offset by 76 ppm to match the weighted-average transit depth of the MIRI observations in order to account for a mismatch in the system parameters applied in analyzing these two data sets and the potential for other epoch-to-epoch changes in the stellar brightness profile.  Models with higher metallicity and thicker haze provide a qualitatively better fit to the transmission spectrum, in line with our findings from the thermal emission data.  A more detailed interpretation of the MIRI transmission spectrum will be presented in Gao et al.~(submitted). All error bars are 1$\sigma$.}
    \label{fig:transmission} 
\end{figure}

%\begin{figure}[h]
%    \centering
%    \includegraphics[width=0.9\textwidth]{gcm_initial_pt_profiles.pdf}
%    \caption{\textbf{Initial pressure-temperature profiles used in the GCM.} For the $300 \times$ solar simulation, the initial profile of the $100 \times$ solar case was used.}
%    \label{fig:gcm_initial_pt_profiles} 
%\end{figure}

\begin{table}[h!]
    \footnotesize
    \caption{\sep\textbf{Derived phase curve parameters. All units are ppm. The penultimate column (RMS) gives the standard deviation of the unbinned residuals, while the final column (RMS$_b$) is the standard deviation of the residuals binned to 5 degrees in phase (277 integrations).}}
    \centering
    \setlength{\tabcolsep}{6pt}
    \renewcommand*{\arraystretch}{1.4}
    \begin{tabular}{c C C C C C C C}
        \hline \hline
       $\lambda$ ($\mu$m) & E & C_1 & D_1 & C_2 & D_2 & \mathrm{RMS} & \mathrm{RMS}_b \\  
%        $\lambda$ ($\mu$m) & $E$ & $C_1$ & $D_1$ & $C_2$ & $D_2$ & RMS & RMS$_b$ \\
        
        \hline\\[-3mm]
        5.0 -- 12.0 & 379_{-13}^{+12} & 127_{-4}^{+4} & -139_{-7}^{+7} & 46_{-4}^{+4} & -15_{-4}^{+4} & 280 & 31\\[1mm]
        \hline\\[-3mm]
        5.0 -- 5.5 & 134_{-28}^{+30} & 33_{-33}^{+20} & -54_{-28}^{+29} & 6_{-16}^{+14} & -8_{-9}^{+9} & 667 & 70\\
        5.5 -- 6.0 & 184_{-31}^{+30} & 75_{-22}^{+18} & -68_{-21}^{+21} & 28_{-11}^{+11} & -21_{-9}^{+9} & 703 & 67\\
        6.0 -- 6.5 & 228_{-29}^{+30} & 89_{-22}^{+16} & -45_{-22}^{+23} & -5_{-12}^{+11} & -23_{-9}^{+9} & 673 & 58\\
        6.5 -- 7.0 & 241_{-33}^{+33} & 116_{-18}^{+13} & -146_{-21}^{+22} & 31_{-12}^{+11} & -24_{-10}^{+10} & 765 & 62\\
        7.0 -- 7.5 & 324_{-39}^{+39} & 154_{-19}^{+13} & -45_{-24}^{+29} & 20_{-14}^{+12} & -6_{-12}^{+11} & 803 & 75\\
        7.5 -- 8.0 & 426_{-41}^{+46} & 196_{-11}^{+10} & -164_{-19}^{+23} & 42_{-11}^{+10} & 1_{-12}^{+12} & 879 & 72\\
        8.0 -- 8.5 & 474_{-40}^{+39} & 133_{-14}^{+12} & -237_{-20}^{+23} & 67_{-12}^{+12} & -72_{-12}^{+13} & 961 & 62\\
        8.5 -- 9.0 & 631_{-47}^{+45} & 91_{-34}^{+21} & -90_{-30}^{+35} & 50_{-18}^{+16} & -4_{-14}^{+14} & 1084 & 69\\
        9.0 -- 9.5 & 604_{-49}^{+49} & 103_{-12}^{+12} & -161_{-21}^{+21} & 63_{-13}^{+12} & 31_{-14}^{+14} & 1165 & 86\\
        9.5 -- 10.0 & 758_{-57}^{+57} & 161_{-18}^{+16} & -210_{-30}^{+34} & 52_{-16}^{+16} & -10_{-17}^{+17} & 1268 & 76\\
        10.0 -- 10.5 & 881_{-70}^{+69} & 95_{-17}^{+17} & -281_{-33}^{+35} & 82_{-18}^{+18} & -5_{-20}^{+19} & 1531 & 108\\
        10.5 -- 11.0 & 962_{-83}^{+83} & 183_{-21}^{+21} & -380_{-35}^{+35} & 129_{-21}^{+21} & -40_{-26}^{+26} & 2067 & 156\\
        11.0 -- 11.5 & 1174_{-114}^{+110} & 271_{-30}^{+31} & -275_{-55}^{+54} & 174_{-30}^{+31} & -3_{-35}^{+34} & 2786 & 237\\
        11.5 -- 12.0 & 1180_{-150}^{+148} & 284_{-41}^{+43} & -94_{-72}^{+68} & 218_{-40}^{+40} & 312_{-50}^{+49} & 3848 & 334\\[1mm]
		\hline
		\hline 
    \end{tabular}
    \label{tab:phase_curve_params}
\end{table}

\begin{table}
    \footnotesize
    \caption{\sep\textbf{Transit parameters inferred from white light curve.}}
    \centering
    \setlength{\tabcolsep}{6pt}
    \renewcommand*{\arraystretch}{1.4}
    \begin{tabular}{c c}
        \hline \hline
        Parameter &  Value \\
        \hline
		$T_0$ [BJD (TDB)] & $2459782.0176719 \pm 8.3 \times 10^{-6}$ \\
		$T_e$ [BJD (TDB)] & $2459782.80880 \pm 1.8 \times 10^{-4}$ \\
		$a/R_*$ & $14.927_{-0.067}^{+0.072}$ \\
		$b$ & $0.282_{-0.016}^{+0.014}$ \\
		\hline
		\hline 
    \end{tabular}
    \label{tab:transit_params}
\end{table}

\begin{table}
    \footnotesize
    \caption{\sep\textbf{Overview of GCM simulations}}
    \centering
    \setlength{\tabcolsep}{6pt}
    \renewcommand*{\arraystretch}{1.4}
    \begin{tabular}{c c c}
        \hline \hline
        metallicity &  haze scaling factor & haze optical properties \\
        \hline
		solar & --- & ---  \\
  		solar & 1 & soot  \\
      	solar & 1 & max. refl.  \\
            solar & 10 & max. refl.\\
		$100 \times$ solar & --- & ---  \\
		$100 \times$ solar & 0.1 & soot  \\
  		$100 \times$ solar & 1 & soot  \\
		$100 \times$ solar & 10 & soot  \\
    	$100 \times$ solar & 1 & tholin  \\
		$100 \times$ solar & 10 & tholin  \\
    	$100 \times$ solar & 1 & max. refl.  \\
		$100 \times$ solar & 10 & max. refl.  \\
		$300 \times$ solar & --- & ---  \\
		$300 \times$ solar & 1 & soot  \\
		$300 \times$ solar & 10 & soot  \\
		$300 \times$ solar & 1 & tholin  \\
		$300 \times$ solar & 10 & tholin  \\
		$300 \times$ solar & 1 & max. refl.  \\
		$300 \times$ solar & 10 & max.refl.  \\
		$300 \times$ solar & 100 & max.refl.  \\
		$3,000 \times$ solar & --- & ---  \\ 
		$3,000 \times$ solar & 1 & soot  \\ 
		$3,000 \times$ solar & 10 & max. refl. \\ 
		$3,000 \times$ solar & 100 & max. refl.  \\ 
		\hline
		\hline 
    \end{tabular}
    \label{tab:gcm_overview}
\end{table}

%\begin{table}
%    \footnotesize
%    \caption{\sep\textbf{adopted system parameters}}
%    \centering
%    \setlength{\tabcolsep}{6pt}
%    \renewcommand*{\arraystretch}{1.4}
%    \begin{tabular}{c c c}
%        \hline \hline
%        Parameter &  Value & Source \\
%        \hline
%		stellar radius, $R_{\star}$ [$R_\odot$] & $0.207 \pm 0.012$ & This work \\
%		stellar mass, $M_{\star}$ [$M_\odot$] & $0.178 \pm 0.010$ & \cite{cloutier21} \\
%		stellar effective temperature, $T_{eff}$ [K] & $3250 \pm 100$ & \cite{cloutier21} \\
%    stellar metallicity, [M/H] [dex] & $0.2 \pm 0.3$ & This work \\
%		orbital period, $P$ [days] & $1.58040433 \pm 0.00000013$ & \cite{cloutier21}\\
%            orbital semi-major axis, $a$ [AU] & ??? & This work \\  
%		planet radius, $R_p$ [$R_{\oplus}$] & $ $ & This work \\
%		\hline
%		\hline 
%    \end{tabular}
%    \label{tab:system_params}
%\end{table}

\clearpage

\noindent \textbf{Data Availability}
The raw data from this study will become publicly available via the Space Science Telescope Institute's Mikulski Archive for Space Telescopes (\url{https://archive.stsci.edu/}) on July 20, 2023. 
 The following Zenodo repository hosts secondary data products including the white light and spectral light curves, extracted fit parameters, and ipython notebooks to calculate derived quantities: \url{https://zenodo.org/record/7703086#.ZAZk1dLMJhE}.
\\

\noindent \textbf{Code Availability}\\
The primary data reduction code used in this paper (\texttt{SPARTA}) is available at \url{https://github.com/ideasrule/sparta}.  The  \texttt{Eureka!} code used for ancillary data analysis is available at \url{https://github.com/kevin218/Eureka}.  We used adapted versions of the SPARC/MITgcm (\url{https://github.com/MITgcm/MITgcm}) and \texttt{CARMA} (\url{https://github.com/ESCOMP/CARMA}) for our GCM and 1-D aerosol modeling, respectively.  The 1-D temperature-pressure profiles used to initialize the GCMs were generated by \texttt{HELIOS} (\url{https://github.com/exoclime/HELIOS}). \\ 

%HyDRo... Chimera...\\

\noindent \textbf{Acknowledgements}
\noindent This work is based on observations made with the NASA/ESA/CSA James Webb Space Telescope. The data were obtained from the Mikulski Archive for Space Telescopes at the Space Telescope Science Institute, which is operated by the Association of Universities for Research in Astronomy, Inc., under NASA contract NAS 5-03127 for JWST. These observations are associated with program \#1803. Support for this program was provided by NASA through a grant from the Space Telescope Science Institute.  This work benefited from the 2022 Exoplanet Summer Program in the Other Worlds Laboratory (OWL) at the University of California, Santa Cruz, a program funded by the Heising-Simons Foundation. EMRK acknowledges funding from the NSF CAREER program (grant \#1931736).  MZ acknowledges support from the 51 Pegasi b Fellowship funded by the Heising-Simons Foundation.  M. Mansfield and LW acknowledge support provided by NASA through the NASA Hubble Fellowship Program. JT acknowledges support from the John Fell Fund and the Candadian Space Agency.\\

\noindent \textbf{Author Contributions}\\
EMRK and JLB proposed for the observations and co-led the project.
EMRK led the writing of the paper.
JLB planned the observations and managed the data analysis.
MZ performed the primary data reduction.  
MES, IM, MTR, VP, ER, ABS, KEA, and TK ran, post-processed, and analyzed GCMs.  
AAAP, JT, MCN, JI, LW, and PM performed retrieval analyses.  
PG calculated 1-D haze profiles and provided expertise on aerosol physics. 
M.\,Malik performed \mbox{1-D} forward models of GJ\,1214b.
QX inverted the observations to generate the global temperature map shown in Figure~\ref{fig:T_map}. 
KBS, TJB, SZ, ED, AD, and POL performed supplementary data reductions.  
KBS, M.\,Mansfield, and GF aided in planning the observing strategy.  
SK provided expertise on the MIRI instrument.
KGS and TB characterized the star.
GWH performed photometric monitoring of the star.
RL provided opacity tables for high mean molecular weight atmosphere modeling.\\

\noindent \textbf{Competing Interests}\\
\noindent The authors declare no competing interests.

%%===========================================================================================%%
%% If you are submitting to one of the Nature Portfolio journals, using the eJP submission   %%
%% system, please include the references within the manuscript file itself. You may do this  %%
%% by copying the reference list from your .bbl file, paste it into the main manuscript .tex %%
%% file, and delete the associated \verb+\bibliography+ commands.                            %%
%%===========================================================================================%%

\bibliography{bibliography}% common bib file

\begin{thebibliography}{100}
\expandafter\ifx\csname url\endcsname\relax
  \def\url#1{\burl{#1}}\fi
\expandafter\ifx\csname urlprefix\endcsname\relax\def\urlprefix{URL }\fi
\providecommand{\bibinfo}[2]{#2}
\providecommand{\eprint}[2][]{\url{#2}}
\providecommand{\doi}[1]{\url{https://doi.org/#1}}
\bibcommenthead

\bibitem{howard12}
\bibinfo{author}{{Howard}, A.~W.} \emph{et~al.}
\newblock \bibinfo{title}{{Planet Occurrence within 0.25 AU of Solar-type Stars
  from Kepler}}.
\newblock \emph{\bibinfo{journal}{\apjs}} \textbf{\bibinfo{volume}{201}}~(2),
  \bibinfo{pages}{15} (\bibinfo{year}{2012}).
\newblock \doi{10.1088/0067-0049/201/2/15},
  \bibinfo{eprint}{{\href{https://arxiv.org/abs/1103.2541}{{arXiv:1103.2541}}}}
   {[astro-ph.EP]}.

\bibitem{fulton17}
\bibinfo{author}{{Fulton}, B.~J.} \emph{et~al.}
\newblock \bibinfo{title}{{The California-Kepler Survey. III. A Gap in the
  Radius Distribution of Small Planets}}.
\newblock \emph{\bibinfo{journal}{\aj}} \textbf{\bibinfo{volume}{154}}~(3),
  \bibinfo{pages}{109} (\bibinfo{year}{2017}).
\newblock \doi{10.3847/1538-3881/aa80eb},
  \bibinfo{eprint}{{\href{https://arxiv.org/abs/1703.10375}{{arXiv:1703.10375}}}}
   {[astro-ph.EP]}.

\bibitem{vaneylen18}
\bibinfo{author}{{Van Eylen}, V.} \emph{et~al.}
\newblock \bibinfo{title}{{An asteroseismic view of the radius valley: stripped
  cores, not born rocky}}.
\newblock \emph{\bibinfo{journal}{\mnras}} \textbf{\bibinfo{volume}{479}}~(4),
  \bibinfo{pages}{4786--4795} (\bibinfo{year}{2018}).
\newblock \doi{10.1093/mnras/sty1783},
  \bibinfo{eprint}{{\href{https://arxiv.org/abs/1710.05398}{{arXiv:1710.05398}}}}
   {[astro-ph.EP]}.

\bibitem{bean21}
\bibinfo{author}{{Bean}, J.~L.}, \bibinfo{author}{{Raymond}, S.~N.} \&
  \bibinfo{author}{{Owen}, J.~E.}
\newblock \bibinfo{title}{{The Nature and Origins of Sub-Neptune Size
  Planets}}.
\newblock \emph{\bibinfo{journal}{Journal of Geophysical Research (Planets)}}
  \textbf{\bibinfo{volume}{126}}~(1), \bibinfo{pages}{e06639}
  (\bibinfo{year}{2021}).
\newblock \doi{10.1029/2020JE006639},
  \bibinfo{eprint}{{\href{https://arxiv.org/abs/2010.11867}{{arXiv:2010.11867}}}}
   {[astro-ph.EP]}.

\bibitem{bean10}
\bibinfo{author}{{Bean}, J.~L.}, \bibinfo{author}{{Miller-Ricci Kempton}, E.}
  \& \bibinfo{author}{{Homeier}, D.}
\newblock \bibinfo{title}{{A ground-based transmission spectrum of the
  super-Earth exoplanet GJ 1214b}}.
\newblock \emph{\bibinfo{journal}{\nat}} \textbf{\bibinfo{volume}{468}}~(7324),
  \bibinfo{pages}{669--672} (\bibinfo{year}{2010}).
\newblock \doi{10.1038/nature09596},
  \bibinfo{eprint}{{\href{https://arxiv.org/abs/1012.0331}{{arXiv:1012.0331}}}}
   {[astro-ph.EP]}.

\bibitem{croll11}
\bibinfo{author}{{Croll}, B.} \emph{et~al.}
\newblock \bibinfo{title}{{Broadband Transmission Spectroscopy of the
  Super-Earth GJ 1214b Suggests a Low Mean Molecular Weight Atmosphere}}.
\newblock \emph{\bibinfo{journal}{\apj}} \textbf{\bibinfo{volume}{736}}~(2),
  \bibinfo{pages}{78} (\bibinfo{year}{2011}).
\newblock \doi{10.1088/0004-637X/736/2/78},
  \bibinfo{eprint}{{\href{https://arxiv.org/abs/1104.0011}{{arXiv:1104.0011}}}}
   {[astro-ph.EP]}.

\bibitem{bean11}
\bibinfo{author}{{Bean}, J.~L.} \emph{et~al.}
\newblock \bibinfo{title}{{The Optical and Near-infrared Transmission Spectrum
  of the Super-Earth GJ 1214b: Further Evidence for a Metal-rich Atmosphere}}.
\newblock \emph{\bibinfo{journal}{\apj}} \textbf{\bibinfo{volume}{743}}~(1),
  \bibinfo{pages}{92} (\bibinfo{year}{2011}).
\newblock \doi{10.1088/0004-637X/743/1/92},
  \bibinfo{eprint}{{\href{https://arxiv.org/abs/1109.0582}{{arXiv:1109.0582}}}}
   {[astro-ph.EP]}.

\bibitem{desert11}
\bibinfo{author}{{D{\'e}sert}, J.-M.} \emph{et~al.}
\newblock \bibinfo{title}{{Observational Evidence for a Metal-rich Atmosphere
  on the Super-Earth GJ1214b}}.
\newblock \emph{\bibinfo{journal}{\apjl}} \textbf{\bibinfo{volume}{731}}~(2),
  \bibinfo{pages}{L40} (\bibinfo{year}{2011}).
\newblock \doi{10.1088/2041-8205/731/2/L40},
  \bibinfo{eprint}{{\href{https://arxiv.org/abs/1103.2370}{{arXiv:1103.2370}}}}
   {[astro-ph.EP]}.

\bibitem{berta12}
\bibinfo{author}{{Berta}, Z.~K.} \emph{et~al.}
\newblock \bibinfo{title}{{The Flat Transmission Spectrum of the Super-Earth
  GJ1214b from Wide Field Camera 3 on the Hubble Space Telescope}}.
\newblock \emph{\bibinfo{journal}{\apj}} \textbf{\bibinfo{volume}{747}}~(1),
  \bibinfo{pages}{35} (\bibinfo{year}{2012}).
\newblock \doi{10.1088/0004-637X/747/1/35},
  \bibinfo{eprint}{{\href{https://arxiv.org/abs/1111.5621}{{arXiv:1111.5621}}}}
   {[astro-ph.EP]}.

\bibitem{fraine13}
\bibinfo{author}{{Fraine}, J.~D.} \emph{et~al.}
\newblock \bibinfo{title}{{Spitzer Transits of the Super-Earth GJ1214b and
  Implications for its Atmosphere}}.
\newblock \emph{\bibinfo{journal}{\apj}} \textbf{\bibinfo{volume}{765}}~(2),
  \bibinfo{pages}{127} (\bibinfo{year}{2013}).
\newblock \doi{10.1088/0004-637X/765/2/127},
  \bibinfo{eprint}{{\href{https://arxiv.org/abs/1301.6763}{{arXiv:1301.6763}}}}
   {[astro-ph.EP]}.

\bibitem{kreidberg14}
\bibinfo{author}{{Kreidberg}, L.} \emph{et~al.}
\newblock \bibinfo{title}{{Clouds in the atmosphere of the super-Earth
  exoplanet GJ1214b}}.
\newblock \emph{\bibinfo{journal}{\nat}} \textbf{\bibinfo{volume}{505}}~(7481),
  \bibinfo{pages}{69--72} (\bibinfo{year}{2014}).
\newblock \doi{10.1038/nature12888},
  \bibinfo{eprint}{{\href{https://arxiv.org/abs/1401.0022}{{arXiv:1401.0022}}}}
   {[astro-ph.EP]}.

\bibitem{kasper20}
\bibinfo{author}{{Kasper}, D.} \emph{et~al.}
\newblock \bibinfo{title}{{Nondetection of Helium in the Upper Atmospheres of
  Three Sub-Neptune Exoplanets}}.
\newblock \emph{\bibinfo{journal}{\aj}} \textbf{\bibinfo{volume}{160}}~(6),
  \bibinfo{pages}{258} (\bibinfo{year}{2020}).
\newblock \doi{10.3847/1538-3881/abbee6},
  \bibinfo{eprint}{{\href{https://arxiv.org/abs/2007.12968}{{arXiv:2007.12968}}}}
   {[astro-ph.EP]}.

\bibitem{orell22}
\bibinfo{author}{{Orell-Miquel}, J.} \emph{et~al.}
\newblock \bibinfo{title}{{A tentative detection of He I in the atmosphere of
  GJ 1214 b}}.
\newblock \emph{\bibinfo{journal}{\aap}} \textbf{\bibinfo{volume}{659}},
  \bibinfo{pages}{A55} (\bibinfo{year}{2022}).
\newblock \doi{10.1051/0004-6361/202142455},
  \bibinfo{eprint}{{\href{https://arxiv.org/abs/2201.11120}{{arXiv:2201.11120}}}}
   {[astro-ph.EP]}.

\bibitem{spake22}
\bibinfo{author}{{Spake}, J.~J.} \emph{et~al.}
\newblock \bibinfo{title}{{Non-detection of He I in the Atmosphere of GJ 1214b
  with Keck/NIRSPEC, at a Time of Minimal Telluric Contamination}}.
\newblock \emph{\bibinfo{journal}{\apjl}} \textbf{\bibinfo{volume}{939}}~(1),
  \bibinfo{pages}{L11} (\bibinfo{year}{2022}).
\newblock \doi{10.3847/2041-8213/ac88c9},
  \bibinfo{eprint}{{\href{https://arxiv.org/abs/2209.03502}{{arXiv:2209.03502}}}}
   {[astro-ph.EP]}.

\bibitem{charbonneau09}
\bibinfo{author}{{Charbonneau}, D.} \emph{et~al.}
\newblock \bibinfo{title}{{A super-Earth transiting a nearby low-mass star}}.
\newblock \emph{\bibinfo{journal}{\nat}} \textbf{\bibinfo{volume}{462}}~(7275),
  \bibinfo{pages}{891--894} (\bibinfo{year}{2009}).
\newblock \doi{10.1038/nature08679},
  \bibinfo{eprint}{{\href{https://arxiv.org/abs/0912.3229}{{arXiv:0912.3229}}}}
   {[astro-ph.EP]}.

\bibitem{kendrew15}
\bibinfo{author}{{Kendrew}, S.} \emph{et~al.}
\newblock \bibinfo{title}{{The Mid-Infrared Instrument for the James Webb Space
  Telescope, IV: The Low-Resolution Spectrometer}}.
\newblock \emph{\bibinfo{journal}{\pasp}} \textbf{\bibinfo{volume}{127}}~(953),
  \bibinfo{pages}{623} (\bibinfo{year}{2015}).
\newblock \doi{10.1086/682255},
  \bibinfo{eprint}{{\href{https://arxiv.org/abs/1512.03000}{{arXiv:1512.03000}}}}
   {[astro-ph.IM]}.

\bibitem{gillon14}
\bibinfo{author}{{Gillon}, M.} \emph{et~al.}
\newblock \bibinfo{title}{{Search for a habitable terrestrial planet transiting
  the nearby red dwarf GJ 1214}}.
\newblock \emph{\bibinfo{journal}{\aap}} \textbf{\bibinfo{volume}{563}},
  \bibinfo{pages}{A21} (\bibinfo{year}{2014}).
\newblock \doi{10.1051/0004-6361/201322362},
  \bibinfo{eprint}{{\href{https://arxiv.org/abs/1307.6722}{{arXiv:1307.6722}}}}
   {[astro-ph.EP]}.

\bibitem{cloutier21}
\bibinfo{author}{{Cloutier}, R.}, \bibinfo{author}{{Charbonneau}, D.},
  \bibinfo{author}{{Deming}, D.}, \bibinfo{author}{{Bonfils}, X.} \&
  \bibinfo{author}{{Astudillo-Defru}, N.}
\newblock \bibinfo{title}{{A More Precise Mass for GJ 1214 b and the Frequency
  of Multiplanet Systems Around Mid-M Dwarfs}}.
\newblock \emph{\bibinfo{journal}{\aj}} \textbf{\bibinfo{volume}{162}}~(5),
  \bibinfo{pages}{174} (\bibinfo{year}{2021}).
\newblock \doi{10.3847/1538-3881/ac1584},
  \bibinfo{eprint}{{\href{https://arxiv.org/abs/2107.14732}{{arXiv:2107.14732}}}}
   {[astro-ph.EP]}.

\bibitem{rowe08}
\bibinfo{author}{{Rowe}, J.~F.} \emph{et~al.}
\newblock \bibinfo{title}{{The Very Low Albedo of an Extrasolar Planet: MOST
  Space-based Photometry of HD 209458}}.
\newblock \emph{\bibinfo{journal}{\apj}} \textbf{\bibinfo{volume}{689}}~(2),
  \bibinfo{pages}{1345--1353} (\bibinfo{year}{2008}).
\newblock \doi{10.1086/591835},
  \bibinfo{eprint}{{\href{https://arxiv.org/abs/0711.4111}{{arXiv:0711.4111}}}}
   {[astro-ph]}.

\bibitem{stevenson14}
\bibinfo{author}{{Stevenson}, K.~B.} \emph{et~al.}
\newblock \bibinfo{title}{{Thermal structure of an exoplanet atmosphere from
  phase-resolved emission spectroscopy}}.
\newblock \emph{\bibinfo{journal}{Science}}
  \textbf{\bibinfo{volume}{346}}~(6211), \bibinfo{pages}{838--841}
  (\bibinfo{year}{2014}).
\newblock \doi{10.1126/science.1256758},
  \bibinfo{eprint}{{\href{https://arxiv.org/abs/1410.2241}{{arXiv:1410.2241}}}}
   {[astro-ph.EP]}.

\bibitem{brandeker22}
\bibinfo{author}{{Brandeker}, A.} \emph{et~al.}
\newblock \bibinfo{title}{{CHEOPS geometric albedo of the hot Jupiter HD 209458
  b}}.
\newblock \emph{\bibinfo{journal}{\aap}} \textbf{\bibinfo{volume}{659}},
  \bibinfo{pages}{L4} (\bibinfo{year}{2022}).
\newblock \doi{10.1051/0004-6361/202243082},
  \bibinfo{eprint}{{\href{https://arxiv.org/abs/2202.11516}{{arXiv:2202.11516}}}}
   {[astro-ph.EP]}.

\bibitem{moroz81}
\bibinfo{author}{{Moroz}, V.~I.}
\newblock \bibinfo{title}{{The Atmosphere of Venus}}.
\newblock \emph{\bibinfo{journal}{Space Science Reviews}}
  \textbf{\bibinfo{volume}{29}}~(1), \bibinfo{pages}{3--127}
  (\bibinfo{year}{1981}).
\newblock \doi{10.1007/BF00177144} .

\bibitem{li18}
\bibinfo{author}{{Li}, L.} \emph{et~al.}
\newblock \bibinfo{title}{{Less absorbed solar energy and more internal heat
  for Jupiter}}.
\newblock \emph{\bibinfo{journal}{Nature Communications}}
  \textbf{\bibinfo{volume}{9}}, \bibinfo{pages}{3709} (\bibinfo{year}{2018}).
\newblock \doi{10.1038/s41467-018-06107-2} .

\bibitem{keating17}
\bibinfo{author}{{Keating}, D.} \& \bibinfo{author}{{Cowan}, N.~B.}
\newblock \bibinfo{title}{{Revisiting the Energy Budget of WASP-43b: Enhanced
  Day-Night Heat Transport}}.
\newblock \emph{\bibinfo{journal}{\apjl}} \textbf{\bibinfo{volume}{849}}~(1),
  \bibinfo{pages}{L5} (\bibinfo{year}{2017}).
\newblock \doi{10.3847/2041-8213/aa8b6b},
  \bibinfo{eprint}{{\href{https://arxiv.org/abs/1709.03502}{{arXiv:1709.03502}}}}
   {[astro-ph.EP]}.

\bibitem{morley15}
\bibinfo{author}{{Morley}, C.~V.} \emph{et~al.}
\newblock \bibinfo{title}{{Thermal Emission and Reflected Light Spectra of
  Super Earths with Flat Transmission Spectra}}.
\newblock \emph{\bibinfo{journal}{\apj}} \textbf{\bibinfo{volume}{815}}~(2),
  \bibinfo{pages}{110} (\bibinfo{year}{2015}).
\newblock \doi{10.1088/0004-637X/815/2/110},
  \bibinfo{eprint}{{\href{https://arxiv.org/abs/1511.01492}{{arXiv:1511.01492}}}}
   {[astro-ph.EP]}.

\bibitem{kawashima19b}
\bibinfo{author}{{Kawashima}, Y.} \& \bibinfo{author}{{Ikoma}, M.}
\newblock \bibinfo{title}{{Theoretical Transmission Spectra of Exoplanet
  Atmospheres with Hydrocarbon Haze: Effect of Creation, Growth, and Settling
  of Haze Particles. II. Dependence on UV Irradiation Intensity, Metallicity,
  C/O Ratio, Eddy Diffusion Coefficient, and Temperature}}.
\newblock \emph{\bibinfo{journal}{\apj}} \textbf{\bibinfo{volume}{877}}~(2),
  \bibinfo{pages}{109} (\bibinfo{year}{2019}).
\newblock \doi{10.3847/1538-4357/ab1b1d} .

\bibitem{adams19}
\bibinfo{author}{{Adams}, D.}, \bibinfo{author}{{Gao}, P.},
  \bibinfo{author}{{de Pater}, I.} \& \bibinfo{author}{{Morley}, C.~V.}
\newblock \bibinfo{title}{{Aggregate Hazes in Exoplanet Atmospheres}}.
\newblock \emph{\bibinfo{journal}{\apj}} \textbf{\bibinfo{volume}{874}}~(1),
  \bibinfo{pages}{61} (\bibinfo{year}{2019}).
\newblock \doi{10.3847/1538-4357/ab074c},
  \bibinfo{eprint}{{\href{https://arxiv.org/abs/1902.05231}{{arXiv:1902.05231}}}}
   {[astro-ph.EP]}.

\bibitem{lavvas19}
\bibinfo{author}{{Lavvas}, P.}, \bibinfo{author}{{Koskinen}, T.},
  \bibinfo{author}{{Steinrueck}, M.~E.}, \bibinfo{author}{{Garc{\'\i}a
  Mu{\~n}oz}, A.} \& \bibinfo{author}{{Showman}, A.~P.}
\newblock \bibinfo{title}{{Photochemical Hazes in Sub-Neptunian Atmospheres
  with a Focus on GJ 1214b}}.
\newblock \emph{\bibinfo{journal}{\apj}} \textbf{\bibinfo{volume}{878}}~(2),
  \bibinfo{pages}{118} (\bibinfo{year}{2019}).
\newblock \doi{10.3847/1538-4357/ab204e},
  \bibinfo{eprint}{{\href{https://arxiv.org/abs/1905.02976}{{arXiv:1905.02976}}}}
   {[astro-ph.EP]}.

\bibitem{gao20b}
\bibinfo{author}{{Gao}, P.} \emph{et~al.}
\newblock \bibinfo{title}{{Aerosol composition of hot giant exoplanets
  dominated by silicates and hydrocarbon hazes}}.
\newblock \emph{\bibinfo{journal}{Nature Astronomy}}
  \textbf{\bibinfo{volume}{4}}, \bibinfo{pages}{951--956}
  (\bibinfo{year}{2020}).
\newblock \doi{10.1038/s41550-020-1114-3},
  \bibinfo{eprint}{{\href{https://arxiv.org/abs/2005.11939}{{arXiv:2005.11939}}}}
   {[astro-ph.EP]}.

\bibitem{kataria14}
\bibinfo{author}{{Kataria}, T.}, \bibinfo{author}{{Showman}, A.~P.},
  \bibinfo{author}{{Fortney}, J.~J.}, \bibinfo{author}{{Marley}, M.~S.} \&
  \bibinfo{author}{{Freedman}, R.~S.}
\newblock \bibinfo{title}{{The Atmospheric Circulation of the Super Earth GJ
  1214b: Dependence on Composition and Metallicity}}.
\newblock \emph{\bibinfo{journal}{\apj}} \textbf{\bibinfo{volume}{785}}~(2),
  \bibinfo{pages}{92} (\bibinfo{year}{2014}).
\newblock \doi{10.1088/0004-637X/785/2/92},
  \bibinfo{eprint}{{\href{https://arxiv.org/abs/1401.1898}{{arXiv:1401.1898}}}}
   {[astro-ph.EP]}.

\bibitem{charnay15b}
\bibinfo{author}{{Charnay}, B.}, \bibinfo{author}{{Meadows}, V.} \&
  \bibinfo{author}{{Leconte}, J.}
\newblock \bibinfo{title}{{3D Modeling of GJ1214b's Atmosphere: Vertical Mixing
  Driven by an Anti-Hadley Circulation}}.
\newblock \emph{\bibinfo{journal}{\apj}} \textbf{\bibinfo{volume}{813}}~(1),
  \bibinfo{pages}{15} (\bibinfo{year}{2015}).
\newblock \doi{10.1088/0004-637X/813/1/15},
  \bibinfo{eprint}{{\href{https://arxiv.org/abs/1509.06814}{{arXiv:1509.06814}}}}
   {[astro-ph.EP]}.

\bibitem{charnay15}
\bibinfo{author}{{Charnay}, B.}, \bibinfo{author}{{Meadows}, V.},
  \bibinfo{author}{{Misra}, A.}, \bibinfo{author}{{Leconte}, J.} \&
  \bibinfo{author}{{Arney}, G.}
\newblock \bibinfo{title}{{3D Modeling of GJ1214b{\textquoteright}s Atmosphere:
  Formation of Inhomogeneous High Clouds and Observational Implications}}.
\newblock \emph{\bibinfo{journal}{\apjl}} \textbf{\bibinfo{volume}{813}}~(1),
  \bibinfo{pages}{L1} (\bibinfo{year}{2015}).
\newblock \doi{10.1088/2041-8205/813/1/L1},
  \bibinfo{eprint}{{\href{https://arxiv.org/abs/1510.01706}{{arXiv:1510.01706}}}}
   {[astro-ph.EP]}.

\bibitem{christie22}
\bibinfo{author}{{Christie}, D.~A.} \emph{et~al.}
\newblock \bibinfo{title}{{The impact of phase equilibrium cloud models on GCM
  simulations of GJ 1214b}}.
\newblock \emph{\bibinfo{journal}{\mnras}} \textbf{\bibinfo{volume}{517}}~(1),
  \bibinfo{pages}{1407--1421} (\bibinfo{year}{2022}).
\newblock \doi{10.1093/mnras/stac2763},
  \bibinfo{eprint}{{\href{https://arxiv.org/abs/2209.12205}{{arXiv:2209.12205}}}}
   {[astro-ph.EP]}.

\bibitem{lavvas17}
\bibinfo{author}{{Lavvas}, P.} \& \bibinfo{author}{{Koskinen}, T.}
\newblock \bibinfo{title}{{Aerosol Properties of the Atmospheres of Extrasolar
  Giant Planets}}.
\newblock \emph{\bibinfo{journal}{\apj}} \textbf{\bibinfo{volume}{847}}~(1),
  \bibinfo{pages}{32} (\bibinfo{year}{2017}).
\newblock \doi{10.3847/1538-4357/aa88ce} .

\bibitem{toon1979}
\bibinfo{author}{Toon, O.~B.}, \bibinfo{author}{Turco, R.~P.},
  \bibinfo{author}{Hamill, P.}, \bibinfo{author}{Kiang, C.~S.} \&
  \bibinfo{author}{Whitten, R.~C.}
\newblock \bibinfo{title}{{A one-dimensional model describing aerosol formation
  and evolution in the stratosphere: II. Sensitivity studies and comparison
  with observations}}.
\newblock \emph{\bibinfo{journal}{Journal of the Atmospheric Sciences}}
  \textbf{\bibinfo{volume}{36}}, \bibinfo{pages}{718--736}
  (\bibinfo{year}{1979}) .

\bibitem{ackerman1995}
\bibinfo{author}{{Ackerman}, A.~S.}, \bibinfo{author}{{Toon}, O.~B.} \&
  \bibinfo{author}{{Hobbs}, P.~V.}
\newblock \bibinfo{title}{{Numerical modeling of ship tracks produced by
  injections of cloud condensation nuclei into marine stratiform clouds}}.
\newblock \emph{\bibinfo{journal}{Journal of Geophysical Research}}
  \textbf{\bibinfo{volume}{100}}, \bibinfo{pages}{7121--7133}
  (\bibinfo{year}{1995}).
\newblock \doi{10.1029/95JD00026} .

\bibitem{khare84}
\bibinfo{author}{{Khare}, B.~N.} \emph{et~al.}
\newblock \bibinfo{title}{{Optical constants of organic tholins produced in a
  simulated Titanian atmosphere: From soft x-ray to microwave frequencies}}.
\newblock \emph{\bibinfo{journal}{\icarus}} \textbf{\bibinfo{volume}{60}}~(1),
  \bibinfo{pages}{127--137} (\bibinfo{year}{1984}).
\newblock \doi{10.1016/0019-1035(84)90142-8} .

\bibitem{kempton12}
\bibinfo{author}{{Miller-Ricci Kempton}, E.}, \bibinfo{author}{{Zahnle}, K.} \&
  \bibinfo{author}{{Fortney}, J.~J.}
\newblock \bibinfo{title}{{The Atmospheric Chemistry of GJ 1214b:
  Photochemistry and Clouds}}.
\newblock \emph{\bibinfo{journal}{\apj}} \textbf{\bibinfo{volume}{745}}~(1),
  \bibinfo{pages}{3} (\bibinfo{year}{2012}).
\newblock \doi{10.1088/0004-637X/745/1/3},
  \bibinfo{eprint}{{\href{https://arxiv.org/abs/1104.5477}{{arXiv:1104.5477}}}}
   {[astro-ph.EP]}.

\bibitem{owen13}
\bibinfo{author}{{Owen}, J.~E.} \& \bibinfo{author}{{Wu}, Y.}
\newblock \bibinfo{title}{{Kepler Planets: A Tale of Evaporation}}.
\newblock \emph{\bibinfo{journal}{\apj}} \textbf{\bibinfo{volume}{775}}~(2),
  \bibinfo{pages}{105} (\bibinfo{year}{2013}).
\newblock \doi{10.1088/0004-637X/775/2/105},
  \bibinfo{eprint}{{\href{https://arxiv.org/abs/1303.3899}{{arXiv:1303.3899}}}}
   {[astro-ph.EP]}.

\bibitem{gupta19}
\bibinfo{author}{{Gupta}, A.} \& \bibinfo{author}{{Schlichting}, H.~E.}
\newblock \bibinfo{title}{{Sculpting the valley in the radius distribution of
  small exoplanets as a by-product of planet formation: the core-powered
  mass-loss mechanism}}.
\newblock \emph{\bibinfo{journal}{\mnras}} \textbf{\bibinfo{volume}{487}}~(1),
  \bibinfo{pages}{24--33} (\bibinfo{year}{2019}).
\newblock \doi{10.1093/mnras/stz1230},
  \bibinfo{eprint}{{\href{https://arxiv.org/abs/1811.03202}{{arXiv:1811.03202}}}}
   {[astro-ph.EP]}.

\bibitem{kuchner03}
\bibinfo{author}{{Kuchner}, M.~J.}
\newblock \bibinfo{title}{{Volatile-rich Earth-Mass Planets in the Habitable
  Zone}}.
\newblock \emph{\bibinfo{journal}{\apjl}} \textbf{\bibinfo{volume}{596}}~(1),
  \bibinfo{pages}{L105--L108} (\bibinfo{year}{2003}).
\newblock \doi{10.1086/378397},
  \bibinfo{eprint}{{\href{https://arxiv.org/abs/astro-ph/0303186}{{arXiv:astro-ph/0303186}}}}
   {[astro-ph]}.

\bibitem{leger04}
\bibinfo{author}{{L{\'e}ger}, A.} \emph{et~al.}
\newblock \bibinfo{title}{{A new family of planets? ``Ocean-Planets''}}.
\newblock \emph{\bibinfo{journal}{\icarus}} \textbf{\bibinfo{volume}{169}}~(2),
  \bibinfo{pages}{499--504} (\bibinfo{year}{2004}).
\newblock \doi{10.1016/j.icarus.2004.01.001},
  \bibinfo{eprint}{{\href{https://arxiv.org/abs/astro-ph/0308324}{{arXiv:astro-ph/0308324}}}}
   {[astro-ph]}.

\bibitem{rogers10b}
\bibinfo{author}{{Rogers}, L.~A.} \& \bibinfo{author}{{Seager}, S.}
\newblock \bibinfo{title}{{Three Possible Origins for the Gas Layer on GJ
  1214b}}.
\newblock \emph{\bibinfo{journal}{\apj}} \textbf{\bibinfo{volume}{716}}~(2),
  \bibinfo{pages}{1208--1216} (\bibinfo{year}{2010}).
\newblock \doi{10.1088/0004-637X/716/2/1208},
  \bibinfo{eprint}{{\href{https://arxiv.org/abs/0912.3243}{{arXiv:0912.3243}}}}
   {[astro-ph.EP]}.

\bibitem{horst18}
\bibinfo{author}{{H{\"o}rst}, S.~M.} \emph{et~al.}
\newblock \bibinfo{title}{{Haze production rates in super-Earth and
  mini-Neptune atmosphere experiments}}.
\newblock \emph{\bibinfo{journal}{Nature Astronomy}}
  \textbf{\bibinfo{volume}{2}}, \bibinfo{pages}{303--306}
  (\bibinfo{year}{2018}).
\newblock \doi{10.1038/s41550-018-0397-0},
  \bibinfo{eprint}{{\href{https://arxiv.org/abs/1801.06512}{{arXiv:1801.06512}}}}
   {[astro-ph.EP]}.

\bibitem{he18}
\bibinfo{author}{{He}, C.} \emph{et~al.}
\newblock \bibinfo{title}{{Laboratory Simulations of Haze Formation in the
  Atmospheres of Super-Earths and Mini-Neptunes: Particle Color and Size
  Distribution}}.
\newblock \emph{\bibinfo{journal}{\apjl}} \textbf{\bibinfo{volume}{856}}~(1),
  \bibinfo{pages}{L3} (\bibinfo{year}{2018}).
\newblock \doi{10.3847/2041-8213/aab42b},
  \bibinfo{eprint}{{\href{https://arxiv.org/abs/1803.01706}{{arXiv:1803.01706}}}}
   {[astro-ph.EP]}.

\bibitem{gavilan18}
\bibinfo{author}{{Gavilan}, L.}, \bibinfo{author}{{Carrasco}, N.},
  \bibinfo{author}{{Vr{\o}nning Hoffmann}, S.}, \bibinfo{author}{{Jones},
  N.~C.} \& \bibinfo{author}{{Mason}, N.~J.}
\newblock \bibinfo{title}{{Organic Aerosols in Anoxic and Oxic Atmospheres of
  Earth-like Exoplanets: VUV-MIR Spectroscopy of CHON Tholins}}.
\newblock \emph{\bibinfo{journal}{\apj}} \textbf{\bibinfo{volume}{861}}~(2),
  \bibinfo{pages}{110} (\bibinfo{year}{2018}).
\newblock \doi{10.3847/1538-4357/aac8df} .

\bibitem{ohno18}
\bibinfo{author}{{Ohno}, K.} \& \bibinfo{author}{{Okuzumi}, S.}
\newblock \bibinfo{title}{{Microphysical Modeling of Mineral Clouds in GJ1214 b
  and GJ436 b: Predicting Upper Limits on the Cloud-top Height}}.
\newblock \emph{\bibinfo{journal}{\apj}} \textbf{\bibinfo{volume}{859}}~(1),
  \bibinfo{pages}{34} (\bibinfo{year}{2018}).
\newblock \doi{10.3847/1538-4357/aabee3},
  \bibinfo{eprint}{{\href{https://arxiv.org/abs/1804.05708}{{arXiv:1804.05708}}}}
   {[astro-ph.EP]}.

\bibitem{bouwman22}
\bibinfo{author}{{Bouwman}, J.} \emph{et~al.}
\newblock \bibinfo{title}{{Spectroscopic time series performance of the
  Mid-Infrared Instrument on the JWST}}.
\newblock \emph{\bibinfo{journal}{arXiv e-prints}}
  \bibinfo{pages}{arXiv:2211.16123} (\bibinfo{year}{2022}).
\newblock
  \bibinfo{eprint}{{\href{https://arxiv.org/abs/2211.16123}{{arXiv:2211.16123}}}}
   {[astro-ph.IM]}.

\bibitem{eureka}
\bibinfo{author}{{Bell}, T.} \emph{et~al.}
\newblock \bibinfo{title}{{Eureka!: An End-to-End Pipeline for JWST Time-Series
  Observations}}.
\newblock \emph{\bibinfo{journal}{The Journal of Open Source Software}}
  \textbf{\bibinfo{volume}{7}}~(79), \bibinfo{pages}{4503}
  (\bibinfo{year}{2022}).
\newblock \doi{10.21105/joss.04503},
  \bibinfo{eprint}{{\href{https://arxiv.org/abs/2207.03585}{{arXiv:2207.03585}}}}
   {[astro-ph.IM]}.

\bibitem{fixsen_2000}
\bibinfo{author}{{Fixsen}, D.~J.} \emph{et~al.}
\newblock \bibinfo{title}{{Cosmic-Ray Rejection and Readout Efficiency for
  Large-Area Arrays}}.
\newblock \emph{\bibinfo{journal}{\pasp}} \textbf{\bibinfo{volume}{112}}~(776),
  \bibinfo{pages}{1350--1359} (\bibinfo{year}{2000}).
\newblock \doi{10.1086/316626},
  \bibinfo{eprint}{{\href{https://arxiv.org/abs/astro-ph/0005486}{{arXiv:astro-ph/0005486}}}}
   {[astro-ph]}.

\bibitem{hu_1996}
\bibinfo{author}{{Hu}, G.~Y.} \& \bibinfo{author}{{O'Connell}, R.~F.}
\newblock \bibinfo{title}{{Analytical inversion of symmetric tridiagonal
  matrices}}.
\newblock \emph{\bibinfo{journal}{Journal of Physics A Mathematical General}}
  \textbf{\bibinfo{volume}{29}}~(7), \bibinfo{pages}{1511--1513}
  (\bibinfo{year}{1996}).
\newblock \doi{10.1088/0305-4470/29/7/020} .

\bibitem{henry23}
\bibinfo{author}{{Henry}, G.~W.} \& \bibinfo{author}{{Bean}, J.~L.}
\newblock \bibinfo{title}{{C14 Automatic Imaging Telescope Photometry of
  GJ1214}}.
\newblock \emph{\bibinfo{journal}{arXiv e-prints}}
  \bibinfo{pages}{arXiv:2302.07874} (\bibinfo{year}{2023}).
\newblock \doi{10.48550/arXiv.2302.07874},
  \bibinfo{eprint}{{\href{https://arxiv.org/abs/2302.07874}{{arXiv:2302.07874}}}}
   {[astro-ph.EP]}.

\bibitem{foremanmackey13}
\bibinfo{author}{{Foreman-Mackey}, D.}, \bibinfo{author}{{Hogg}, D.~W.},
  \bibinfo{author}{{Lang}, D.} \& \bibinfo{author}{{Goodman}, J.}
\newblock \bibinfo{title}{{emcee: The MCMC Hammer}}.
\newblock \emph{\bibinfo{journal}{\pasp}} \textbf{\bibinfo{volume}{125}}~(925),
  \bibinfo{pages}{306} (\bibinfo{year}{2013}).
\newblock \doi{10.1086/670067},
  \bibinfo{eprint}{{\href{https://arxiv.org/abs/1202.3665}{{arXiv:1202.3665}}}}
   {[astro-ph.IM]}.

\bibitem{batman}
\bibinfo{author}{{Kreidberg}, L.}
\newblock \bibinfo{title}{{batman: BAsic Transit Model cAlculatioN in Python}}.
\newblock \emph{\bibinfo{journal}{\pasp}} \textbf{\bibinfo{volume}{127}}~(957),
  \bibinfo{pages}{1161} (\bibinfo{year}{2015}).
\newblock \doi{10.1086/683602},
  \bibinfo{eprint}{{\href{https://arxiv.org/abs/1507.08285}{{arXiv:1507.08285}}}}
   {[astro-ph.EP]}.

\bibitem{kokori_2022}
\bibinfo{author}{{Kokori}, A.} \emph{et~al.}
\newblock \bibinfo{title}{{ExoClock Project. II. A Large-scale Integrated Study
  with 180 Updated Exoplanet Ephemerides}}.
\newblock \emph{\bibinfo{journal}{\apjs}} \textbf{\bibinfo{volume}{258}}~(2),
  \bibinfo{pages}{40} (\bibinfo{year}{2022}).
\newblock \doi{10.3847/1538-4365/ac3a10},
  \bibinfo{eprint}{{\href{https://arxiv.org/abs/2110.13863}{{arXiv:2110.13863}}}}
   {[astro-ph.EP]}.

\bibitem{argyriou_phd}
\bibinfo{author}{Argyriou, Y.}
\newblock \emph{\bibinfo{title}{Calibration of the MIRI instrument on board the
  James Webb Space Telescope}}.
\newblock Ph.D. thesis, \bibinfo{school}{Ku Leuven Institute of Astronomy}
  (\bibinfo{year}{2021}).

\bibitem{cowan2008}
\bibinfo{author}{{Cowan}, N.~B.} \& \bibinfo{author}{{Agol}, E.}
\newblock \bibinfo{title}{{Inverting Phase Functions to Map Exoplanets}}.
\newblock \emph{\bibinfo{journal}{\apjl}} \textbf{\bibinfo{volume}{678}}~(2),
  \bibinfo{pages}{L129} (\bibinfo{year}{2008}).
\newblock \doi{10.1086/588553},
  \bibinfo{eprint}{{\href{https://arxiv.org/abs/0803.3622}{{arXiv:0803.3622}}}}
   {[astro-ph]}.

\bibitem{keating19}
\bibinfo{author}{{Keating}, D.}, \bibinfo{author}{{Cowan}, N.~B.} \&
  \bibinfo{author}{{Dang}, L.}
\newblock \bibinfo{title}{{Uniformly hot nightside temperatures on short-period
  gas giants}}.
\newblock \emph{\bibinfo{journal}{Nature Astronomy}}
  \textbf{\bibinfo{volume}{3}}, \bibinfo{pages}{1092--1098}
  (\bibinfo{year}{2019}).
\newblock \doi{10.1038/s41550-019-0859-z},
  \bibinfo{eprint}{{\href{https://arxiv.org/abs/1809.00002}{{arXiv:1809.00002}}}}
   {[astro-ph.EP]}.

\bibitem{showman09}
\bibinfo{author}{{Showman}, A.~P.} \emph{et~al.}
\newblock \bibinfo{title}{{Atmospheric Circulation of Hot Jupiters: Coupled
  Radiative-Dynamical General Circulation Model Simulations of HD 189733b and
  HD 209458b}}.
\newblock \emph{\bibinfo{journal}{\apj}} \textbf{\bibinfo{volume}{699}}~(1),
  \bibinfo{pages}{564--584} (\bibinfo{year}{2009}).
\newblock \doi{10.1088/0004-637X/699/1/564},
  \bibinfo{eprint}{{\href{https://arxiv.org/abs/0809.2089}{{arXiv:0809.2089}}}}
   {[astro-ph]}.

\bibitem{kataria13}
\bibinfo{author}{{Kataria}, T.} \emph{et~al.}
\newblock \bibinfo{title}{{Three-dimensional Atmospheric Circulation of Hot
  Jupiters on Highly Eccentric Orbits}}.
\newblock \emph{\bibinfo{journal}{\apj}} \textbf{\bibinfo{volume}{767}}~(1),
  \bibinfo{pages}{76} (\bibinfo{year}{2013}).
\newblock \doi{10.1088/0004-637X/767/1/76},
  \bibinfo{eprint}{{\href{https://arxiv.org/abs/1208.3795}{{arXiv:1208.3795}}}}
   {[astro-ph.EP]}.

\bibitem{adcroft04}
\bibinfo{author}{{Adcroft}, A.}, \bibinfo{author}{{Campin}, J.-M.},
  \bibinfo{author}{{Hill}, C.} \& \bibinfo{author}{{Marshall}, J.}
\newblock \bibinfo{title}{{Implementation of an Atmosphere Ocean General
  Circulation Model on the Expanded Spherical Cube}}.
\newblock \emph{\bibinfo{journal}{Monthly Weather Review}}
  \textbf{\bibinfo{volume}{132}}~(12), \bibinfo{pages}{2845}
  (\bibinfo{year}{2004}).
\newblock \doi{10.1175/MWR2823.1} .

\bibitem{marley1999}
\bibinfo{author}{{Marley}, M.~S.} \& \bibinfo{author}{{McKay}, C.~P.}
\newblock \bibinfo{title}{{Thermal Structure of Uranus' Atmosphere}}.
\newblock \emph{\bibinfo{journal}{\icarus}} \textbf{\bibinfo{volume}{138}}~(2),
  \bibinfo{pages}{268--286} (\bibinfo{year}{1999}).
\newblock \doi{10.1006/icar.1998.6071} .

\bibitem{liu13}
\bibinfo{author}{{Liu}, B.} \& \bibinfo{author}{{Showman}, A.~P.}
\newblock \bibinfo{title}{{Atmospheric Circulation of Hot Jupiters:
  Insensitivity to Initial Conditions}}.
\newblock \emph{\bibinfo{journal}{\apj}} \textbf{\bibinfo{volume}{770}}~(1),
  \bibinfo{pages}{42} (\bibinfo{year}{2013}).
\newblock \doi{10.1088/0004-637X/770/1/42},
  \bibinfo{eprint}{{\href{https://arxiv.org/abs/1208.0126}{{arXiv:1208.0126}}}}
   {[astro-ph.EP]}.

\bibitem{malik17}
\bibinfo{author}{{Malik}, M.} \emph{et~al.}
\newblock \bibinfo{title}{{HELIOS: An Open-source, GPU-accelerated Radiative
  Transfer Code for Self-consistent Exoplanetary Atmospheres}}.
\newblock \emph{\bibinfo{journal}{\aj}} \textbf{\bibinfo{volume}{153}}~(2),
  \bibinfo{pages}{56} (\bibinfo{year}{2017}).
\newblock \doi{10.3847/1538-3881/153/2/56},
  \bibinfo{eprint}{{\href{https://arxiv.org/abs/1606.05474}{{arXiv:1606.05474}}}}
   {[astro-ph.EP]}.

\bibitem{malik19}
\bibinfo{author}{{Malik}, M.} \emph{et~al.}
\newblock \bibinfo{title}{{Self-luminous and Irradiated Exoplanetary
  Atmospheres Explored with HELIOS}}.
\newblock \emph{\bibinfo{journal}{\aj}} \textbf{\bibinfo{volume}{157}}~(5),
  \bibinfo{pages}{170} (\bibinfo{year}{2019}).
\newblock \doi{10.3847/1538-3881/ab1084},
  \bibinfo{eprint}{{\href{https://arxiv.org/abs/1903.06794}{{arXiv:1903.06794}}}}
   {[astro-ph.EP]}.

\bibitem{zhang17}
\bibinfo{author}{{Zhang}, X.} \& \bibinfo{author}{{Showman}, A.~P.}
\newblock \bibinfo{title}{{Effects of Bulk Composition on the Atmospheric
  Dynamics on Close-in Exoplanets}}.
\newblock \emph{\bibinfo{journal}{\apj}} \textbf{\bibinfo{volume}{836}}~(1),
  \bibinfo{pages}{73} (\bibinfo{year}{2017}).
\newblock \doi{10.3847/1538-4357/836/1/73},
  \bibinfo{eprint}{{\href{https://arxiv.org/abs/1607.04260}{{arXiv:1607.04260}}}}
   {[astro-ph.EP]}.

\bibitem{tomasko2009}
\bibinfo{author}{{Tomasko}, M.~G.}, \bibinfo{author}{{Doose}, L.~R.},
  \bibinfo{author}{{Dafoe}, L.~E.} \& \bibinfo{author}{{See}, C.}
\newblock \bibinfo{title}{{Limits on the size of aerosols from measurements of
  linear polarization in Titan{\textquoteright}s atmosphere}}.
\newblock \emph{\bibinfo{journal}{\icarus}} \textbf{\bibinfo{volume}{204}}~(1),
  \bibinfo{pages}{271--283} (\bibinfo{year}{2009}).
\newblock \doi{10.1016/j.icarus.2009.05.034} .

\bibitem{lavvas2010}
\bibinfo{author}{{Lavvas}, P.}, \bibinfo{author}{{Yelle}, R.~V.} \&
  \bibinfo{author}{{Griffith}, C.~A.}
\newblock \bibinfo{title}{{Titan{\textquoteright}s vertical aerosol structure
  at the Huygens landing site: Constraints on particle size, density, charge,
  and refractive index}}.
\newblock \emph{\bibinfo{journal}{\icarus}} \textbf{\bibinfo{volume}{210}}~(2),
  \bibinfo{pages}{832--842} (\bibinfo{year}{2010}).
\newblock \doi{10.1016/j.icarus.2010.07.025} .

\bibitem{gladstone2016}
\bibinfo{author}{{Gladstone}, G.~R.} \emph{et~al.}
\newblock \bibinfo{title}{{The atmosphere of Pluto as observed by New
  Horizons}}.
\newblock \emph{\bibinfo{journal}{Science}}
  \textbf{\bibinfo{volume}{351}}~(6279), \bibinfo{pages}{aad8866}
  (\bibinfo{year}{2016}).
\newblock \doi{10.1126/science.aad8866},
  \bibinfo{eprint}{{\href{https://arxiv.org/abs/1604.05356}{{arXiv:1604.05356}}}}
   {[astro-ph.EP]}.

\bibitem{parmentier16}
\bibinfo{author}{{Parmentier}, V.}, \bibinfo{author}{{Fortney}, J.~J.},
  \bibinfo{author}{{Showman}, A.~P.}, \bibinfo{author}{{Morley}, C.} \&
  \bibinfo{author}{{Marley}, M.~S.}
\newblock \bibinfo{title}{{Transitions in the Cloud Composition of Hot
  Jupiters}}.
\newblock \emph{\bibinfo{journal}{\apj}} \textbf{\bibinfo{volume}{828}}~(1),
  \bibinfo{pages}{22} (\bibinfo{year}{2016}).
\newblock \doi{10.3847/0004-637X/828/1/22},
  \bibinfo{eprint}{{\href{https://arxiv.org/abs/1602.03088}{{arXiv:1602.03088}}}}
   {[astro-ph.EP]}.

\bibitem{kempton2012constraining}
\bibinfo{author}{Kempton, E. M.-R.} \& \bibinfo{author}{Rauscher, E.}
\newblock \bibinfo{title}{Constraining high-speed winds in exoplanet
  atmospheres through observations of anomalous doppler shifts during transit}.
\newblock \emph{\bibinfo{journal}{The Astrophysical Journal}}
  \textbf{\bibinfo{volume}{751}}~(2), \bibinfo{pages}{117}
  (\bibinfo{year}{2012}) .

\bibitem{savel2023diagnosing}
\bibinfo{author}{Savel, A.~B.} \emph{et~al.}
\newblock \bibinfo{title}{Diagnosing limb asymmetries in hot and ultra-hot
  jupiters with high-resolution transmission spectroscopy}.
\newblock \emph{\bibinfo{journal}{arXiv preprint arXiv:2301.01694}}
  (\bibinfo{year}{2023}) .

\bibitem{harada2021signatures}
\bibinfo{author}{Harada, C.~K.} \emph{et~al.}
\newblock \bibinfo{title}{Signatures of clouds in hot jupiter atmospheres:
  Modeled high-resolution emission spectra from 3d general circulation models}.
\newblock \emph{\bibinfo{journal}{The Astrophysical Journal}}
  \textbf{\bibinfo{volume}{909}}~(1), \bibinfo{pages}{85}
  (\bibinfo{year}{2021}) .

\bibitem{piette22}
\bibinfo{author}{{Piette}, A. A.~A.}, \bibinfo{author}{{Madhusudhan}, N.} \&
  \bibinfo{author}{{Mandell}, A.~M.}
\newblock \bibinfo{title}{{HyDRo: atmospheric retrieval of rocky exoplanets in
  thermal emission}}.
\newblock \emph{\bibinfo{journal}{\mnras}} \textbf{\bibinfo{volume}{511}}~(2),
  \bibinfo{pages}{2565--2584} (\bibinfo{year}{2022}).
\newblock \doi{10.1093/mnras/stab3612},
  \bibinfo{eprint}{{\href{https://arxiv.org/abs/2112.05059}{{arXiv:2112.05059}}}}
   {[astro-ph.EP]}.

\bibitem{Line2013}
\bibinfo{author}{{Line}, M.~R.} \emph{et~al.}
\newblock \bibinfo{title}{{A Systematic Retrieval Analysis of Secondary Eclipse
  Spectra. I. A Comparison of Atmospheric Retrieval Techniques}}.
\newblock \emph{\bibinfo{journal}{\apj}} \textbf{\bibinfo{volume}{775}}~(2),
  \bibinfo{pages}{137} (\bibinfo{year}{2013}).
\newblock \doi{10.1088/0004-637X/775/2/137},
  \bibinfo{eprint}{{\href{https://arxiv.org/abs/1304.5561}{{arXiv:1304.5561}}}}
   {[astro-ph.EP]}.

\bibitem{gandhi18}
\bibinfo{author}{{Gandhi}, S.} \& \bibinfo{author}{{Madhusudhan}, N.}
\newblock \bibinfo{title}{{Retrieval of exoplanet emission spectra with
  HyDRA}}.
\newblock \emph{\bibinfo{journal}{\mnras}} \textbf{\bibinfo{volume}{474}},
  \bibinfo{pages}{271--288} (\bibinfo{year}{2018}).
\newblock \doi{10.1093/mnras/stx2748},
  \bibinfo{eprint}{{\href{https://arxiv.org/abs/1710.06433}{{arXiv:1710.06433}}}}
   {[astro-ph.EP]}.

\bibitem{gandhi20}
\bibinfo{author}{Gandhi, S.}, \bibinfo{author}{Madhusudhan, N.} \&
  \bibinfo{author}{Mandell, A.}
\newblock \bibinfo{title}{H- and {Dissociation} in {Ultra}-hot {Jupiters}: {A}
  {Retrieval} {Case} {Study} of {WASP}-18b}.
\newblock \emph{\bibinfo{journal}{The Astronomical Journal}}
  \textbf{\bibinfo{volume}{159}}, \bibinfo{pages}{232} (\bibinfo{year}{2020}).
\newblock
  \urlprefix\url{https://ui.adsabs.harvard.edu/abs/2020AJ....159..232G}.
\newblock \doi{10.3847/1538-3881/ab845e}, \bibinfo{note}{aDS Bibcode:
  2020AJ....159..232G} .

\bibitem{piette20}
\bibinfo{author}{{Piette}, A. A.~A.} \& \bibinfo{author}{{Madhusudhan}, N.}
\newblock \bibinfo{title}{{Considerations for atmospheric retrieval of
  high-precision brown dwarf spectra}}.
\newblock \emph{\bibinfo{journal}{\mnras}} \textbf{\bibinfo{volume}{497}}~(4),
  \bibinfo{pages}{5136--5154} (\bibinfo{year}{2020}).
\newblock \doi{10.1093/mnras/staa2289},
  \bibinfo{eprint}{{\href{https://arxiv.org/abs/2007.15004}{{arXiv:2007.15004}}}}
   {[astro-ph.EP]}.

\bibitem{skilling06}
\bibinfo{author}{Skilling, J.}
\newblock \bibinfo{title}{Nested sampling for general bayesian computation}.
\newblock \emph{\bibinfo{journal}{Bayesian Anal.}}
  \textbf{\bibinfo{volume}{1}}~(4), \bibinfo{pages}{833--859}
  (\bibinfo{year}{2006}).
\newblock \urlprefix\url{https://doi.org/10.1214/06-BA127}.
\newblock \doi{10.1214/06-BA127} .

\bibitem{feroz09}
\bibinfo{author}{{Feroz}, F.}, \bibinfo{author}{{Hobson}, M.~P.} \&
  \bibinfo{author}{{Bridges}, M.}
\newblock \bibinfo{title}{{MULTINEST: an efficient and robust Bayesian
  inference tool for cosmology and particle physics}}.
\newblock \emph{\bibinfo{journal}{\mnras}} \textbf{\bibinfo{volume}{398}},
  \bibinfo{pages}{1601--1614} (\bibinfo{year}{2009}).
\newblock \doi{10.1111/j.1365-2966.2009.14548.x},
  \bibinfo{eprint}{{\href{https://arxiv.org/abs/0809.3437}{{arXiv:0809.3437}}}}
  .

\bibitem{buchner14}
\bibinfo{author}{{Buchner}, J.} \emph{et~al.}
\newblock \bibinfo{title}{{X-ray spectral modelling of the AGN obscuring region
  in the CDFS: Bayesian model selection and catalogue}}.
\newblock \emph{\bibinfo{journal}{Astronomy and Astrophysics}}
  \textbf{\bibinfo{volume}{564}}, \bibinfo{pages}{A125} (\bibinfo{year}{2014}).
\newblock \doi{10.1051/0004-6361/201322971},
  \bibinfo{eprint}{{\href{https://arxiv.org/abs/1402.0004}{{arXiv:1402.0004}}}}
   {[astro-ph.HE]}.

\bibitem{madhusudhan09}
\bibinfo{author}{{Madhusudhan}, N.} \& \bibinfo{author}{{Seager}, S.}
\newblock \bibinfo{title}{{A Temperature and Abundance Retrieval Method for
  Exoplanet Atmospheres}}.
\newblock \emph{\bibinfo{journal}{\apj}} \textbf{\bibinfo{volume}{707}},
  \bibinfo{pages}{24--39} (\bibinfo{year}{2009}).
\newblock \doi{10.1088/0004-637X/707/1/24},
  \bibinfo{eprint}{{\href{https://arxiv.org/abs/0910.1347}{{arXiv:0910.1347}}}}
   {[astro-ph.EP]}.

\bibitem{rothman10}
\bibinfo{author}{{Rothman}, L.~S.} \emph{et~al.}
\newblock \bibinfo{title}{{HITEMP, the high-temperature molecular spectroscopic
  database}}.
\newblock \emph{\bibinfo{journal}{Journal of Quantitative Spectroscopy and
  Radiative Transfer}} \textbf{\bibinfo{volume}{111}},
  \bibinfo{pages}{2139--2150} (\bibinfo{year}{2010}).
\newblock \doi{10.1016/j.jqsrt.2010.05.001} .

\bibitem{yurchenko13}
\bibinfo{author}{{Yurchenko}, S.~N.}, \bibinfo{author}{{Tennyson}, J.},
  \bibinfo{author}{{Barber}, R.~J.} \& \bibinfo{author}{{Thiel}, W.}
\newblock \bibinfo{title}{{Vibrational transition moments of CH$_{4}$ from
  first principles}}.
\newblock \emph{\bibinfo{journal}{Journal of Molecular Spectroscopy}}
  \textbf{\bibinfo{volume}{291}}, \bibinfo{pages}{69--76}
  (\bibinfo{year}{2013}).
\newblock \doi{10.1016/j.jms.2013.05.014},
  \bibinfo{eprint}{{\href{https://arxiv.org/abs/1302.1720}{{arXiv:1302.1720}}}}
   {[astro-ph.SR]}.

\bibitem{yurchenko14}
\bibinfo{author}{{Yurchenko}, S.~N.} \& \bibinfo{author}{{Tennyson}, J.}
\newblock \bibinfo{title}{{ExoMol line lists - IV. The rotation-vibration
  spectrum of methane up to 1500 K}}.
\newblock \emph{\bibinfo{journal}{\mnras}} \textbf{\bibinfo{volume}{440}},
  \bibinfo{pages}{1649--1661} (\bibinfo{year}{2014}).
\newblock \doi{10.1093/mnras/stu326},
  \bibinfo{eprint}{{\href{https://arxiv.org/abs/1401.4852}{{arXiv:1401.4852}}}}
   {[astro-ph.EP]}.

\bibitem{harris06}
\bibinfo{author}{{Harris}, G.~J.}, \bibinfo{author}{{Tennyson}, J.},
  \bibinfo{author}{{Kaminsky}, B.~M.}, \bibinfo{author}{{Pavlenko}, Y.~V.} \&
  \bibinfo{author}{{Jones}, H.~R.~A.}
\newblock \bibinfo{title}{{Improved HCN/HNC linelist, model atmospheres and
  synthetic spectra for WZ Cas}}.
\newblock \emph{\bibinfo{journal}{\mnras}} \textbf{\bibinfo{volume}{367}}~(1),
  \bibinfo{pages}{400--406} (\bibinfo{year}{2006}).
\newblock \doi{10.1111/j.1365-2966.2005.09960.x},
  \bibinfo{eprint}{{\href{https://arxiv.org/abs/astro-ph/0512363}{{arXiv:astro-ph/0512363}}}}
   {[astro-ph]}.

\bibitem{yurchenko11}
\bibinfo{author}{{Yurchenko}, S.~N.}, \bibinfo{author}{{Barber}, R.~J.} \&
  \bibinfo{author}{{Tennyson}, J.}
\newblock \bibinfo{title}{{A variationally computed line list for hot
  NH$_{3}$}}.
\newblock \emph{\bibinfo{journal}{\mnras}} \textbf{\bibinfo{volume}{413}},
  \bibinfo{pages}{1828--1834} (\bibinfo{year}{2011}).
\newblock \doi{10.1111/j.1365-2966.2011.18261.x},
  \bibinfo{eprint}{{\href{https://arxiv.org/abs/1011.1569}{{arXiv:1011.1569}}}}
   {[astro-ph.EP]}.

\bibitem{barklem16}
\bibinfo{author}{{Barklem}, P.~S.} \& \bibinfo{author}{{Collet}, R.}
\newblock \bibinfo{title}{{Partition functions and equilibrium constants for
  diatomic molecules and atoms of astrophysical interest}}.
\newblock \emph{\bibinfo{journal}{\aap}} \textbf{\bibinfo{volume}{588}},
  \bibinfo{pages}{A96} (\bibinfo{year}{2016}).
\newblock \doi{10.1051/0004-6361/201526961},
  \bibinfo{eprint}{{\href{https://arxiv.org/abs/1602.03304}{{arXiv:1602.03304}}}}
   {[astro-ph.IM]}.

\bibitem{western18}
\bibinfo{author}{{Western}, C.~M.} \emph{et~al.}
\newblock \bibinfo{title}{{The spectrum of N$_{2}$ from 4,500 to 15,700
  cm$^{-1}$ revisited with PGOPHER}}.
\newblock \emph{\bibinfo{journal}{Journal of Quantitative Spectroscopy and
  Radiative Transfer}} \textbf{\bibinfo{volume}{219}},
  \bibinfo{pages}{127--141} (\bibinfo{year}{2018}).
\newblock \doi{10.1016/j.jqsrt.2018.07.017},
  \bibinfo{eprint}{{\href{https://arxiv.org/abs/1905.02528}{{arXiv:1905.02528}}}}
   {[physics.chem-ph]}.

\bibitem{richard12}
\bibinfo{author}{Richard, C.} \emph{et~al.}
\newblock \bibinfo{title}{New section of the hitran database: Collision-induced
  absorption (cia)}.
\newblock \emph{\bibinfo{journal}{Journal of Quantitative Spectroscopy and
  Radiative Transfer}} \textbf{\bibinfo{volume}{113}}~(11),
  \bibinfo{pages}{1276 -- 1285} (\bibinfo{year}{2012}).
\newblock
  \urlprefix\url{http://www.sciencedirect.com/science/article/pii/S0022407311003773}.
\newblock \doi{https://doi.org/10.1016/j.jqsrt.2011.11.004},
  \bibinfo{note}{three Leaders in Spectroscopy} .

\bibitem{gandhi17}
\bibinfo{author}{{Gandhi}, S.} \& \bibinfo{author}{{Madhusudhan}, N.}
\newblock \bibinfo{title}{{genesis: new self-consistent models of exoplanetary
  spectra}}.
\newblock \emph{\bibinfo{journal}{\mnras}} \textbf{\bibinfo{volume}{472}},
  \bibinfo{pages}{2334--2355} (\bibinfo{year}{2017}).
\newblock \doi{10.1093/mnras/stx1601},
  \bibinfo{eprint}{{\href{https://arxiv.org/abs/1706.02302}{{arXiv:1706.02302}}}}
   {[astro-ph.EP]}.

\bibitem{benneke12}
\bibinfo{author}{{Benneke}, B.} \& \bibinfo{author}{{Seager}, S.}
\newblock \bibinfo{title}{{Atmospheric Retrieval for Super-Earths: Uniquely
  Constraining the Atmospheric Composition with Transmission Spectroscopy}}.
\newblock \emph{\bibinfo{journal}{\apj}} \textbf{\bibinfo{volume}{753}},
  \bibinfo{pages}{100} (\bibinfo{year}{2012}).
\newblock \doi{10.1088/0004-637X/753/2/100},
  \bibinfo{eprint}{{\href{https://arxiv.org/abs/1203.4018}{{arXiv:1203.4018}}}}
   {[astro-ph.EP]}.

\bibitem{pinhas17}
\bibinfo{author}{{Pinhas}, A.} \& \bibinfo{author}{{Madhusudhan}, N.}
\newblock \bibinfo{title}{{On signatures of clouds in exoplanetary transit
  spectra}}.
\newblock \emph{\bibinfo{journal}{\mnras}} \textbf{\bibinfo{volume}{471}}~(4),
  \bibinfo{pages}{4355--4373} (\bibinfo{year}{2017}).
\newblock \doi{10.1093/mnras/stx1849},
  \bibinfo{eprint}{{\href{https://arxiv.org/abs/1705.08893}{{arXiv:1705.08893}}}}
   {[astro-ph.EP]}.

\bibitem{trotta08}
\bibinfo{author}{{Trotta}, R.}
\newblock \bibinfo{title}{{Bayes in the sky: Bayesian inference and model
  selection in cosmology}}.
\newblock \emph{\bibinfo{journal}{Contemporary Physics}}
  \textbf{\bibinfo{volume}{49}}~(2), \bibinfo{pages}{71--104}
  (\bibinfo{year}{2008}).
\newblock \doi{10.1080/00107510802066753},
  \bibinfo{eprint}{{\href{https://arxiv.org/abs/0803.4089}{{arXiv:0803.4089}}}}
   {[astro-ph]}.

\bibitem{benneke13}
\bibinfo{author}{Benneke, B.} \& \bibinfo{author}{Seager, S.}
\newblock \bibinfo{title}{{HOW} {TO} {DISTINGUISH} {BETWEEN} {CLOUDY}
  {MINI}-{NEPTUNES} {AND} {WATER}/{VOLATILE}-{DOMINATED} {SUPER}-{EARTHS}}.
\newblock \emph{\bibinfo{journal}{The Astrophysical Journal}}
  \textbf{\bibinfo{volume}{778}}~(2), \bibinfo{pages}{153}
  (\bibinfo{year}{2013}).
\newblock \urlprefix\url{https://doi.org/10.1088/0004-637x/778/2/153}.
\newblock \doi{10.1088/0004-637X/778/2/153}, \bibinfo{note}{publisher: American
  Astronomical Society} .

\bibitem{piette20_MN}
\bibinfo{author}{{Piette}, A. A.~A.} \& \bibinfo{author}{{Madhusudhan}, N.}
\newblock \bibinfo{title}{{On the Temperature Profiles and Emission Spectra of
  Mini-Neptune Atmospheres}}.
\newblock \emph{\bibinfo{journal}{\apj}} \textbf{\bibinfo{volume}{904}}~(2),
  \bibinfo{pages}{154} (\bibinfo{year}{2020}).
\newblock \doi{10.3847/1538-4357/abbfb1},
  \bibinfo{eprint}{{\href{https://arxiv.org/abs/2009.11290}{{arXiv:2009.11290}}}}
   {[astro-ph.EP]}.

\bibitem{welbanks2022_loo}
\bibinfo{author}{{Welbanks}, L.}, \bibinfo{author}{{McGill}, P.},
  \bibinfo{author}{{Line}, M.} \& \bibinfo{author}{{Madhusudhan}, N.}
\newblock \bibinfo{title}{{On the Application of Bayesian Leave-one-out
  Cross-validation to Exoplanet Atmospheric Analysis}}.
\newblock \emph{\bibinfo{journal}{\aj}} \textbf{\bibinfo{volume}{165}}~(3),
  \bibinfo{pages}{112} (\bibinfo{year}{2023}).
\newblock \doi{10.3847/1538-3881/acab67},
  \bibinfo{eprint}{{\href{https://arxiv.org/abs/2212.03872}{{arXiv:2212.03872}}}}
   {[astro-ph.EP]}.

\bibitem{Vehtari2017}
\bibinfo{author}{Vehtari, A.}, \bibinfo{author}{Gelman, A.} \&
  \bibinfo{author}{Gabry, J.}
\newblock \bibinfo{title}{Practical bayesian model evaluation using
  leave-one-out cross-validation and waic}.
\newblock \emph{\bibinfo{journal}{Statistics and computing}}
  \textbf{\bibinfo{volume}{27}}~(5), \bibinfo{pages}{1413--1432}
  (\bibinfo{year}{2017}) .

\bibitem{Barstow2020}
\bibinfo{author}{{Barstow}, J.~K.} \emph{et~al.}
\newblock \bibinfo{title}{{A comparison of exoplanet spectroscopic retrieval
  tools}}.
\newblock \emph{\bibinfo{journal}{\mnras}} \textbf{\bibinfo{volume}{493}}~(4),
  \bibinfo{pages}{4884--4909} (\bibinfo{year}{2020}).
\newblock \doi{10.1093/mnras/staa548},
  \bibinfo{eprint}{{\href{https://arxiv.org/abs/2002.01063}{{arXiv:2002.01063}}}}
   {[astro-ph.EP]}.

\bibitem{Parmentier2014}
\bibinfo{author}{{Parmentier}, V.} \& \bibinfo{author}{{Guillot}, T.}
\newblock \bibinfo{title}{{A non-grey analytical model for irradiated
  atmospheres. I. Derivation}}.
\newblock \emph{\bibinfo{journal}{\aap}} \textbf{\bibinfo{volume}{562}},
  \bibinfo{pages}{A133} (\bibinfo{year}{2014}).
\newblock \doi{10.1051/0004-6361/201322342},
  \bibinfo{eprint}{{\href{https://arxiv.org/abs/1311.6597}{{arXiv:1311.6597}}}}
   {[astro-ph.EP]}.

\end{thebibliography}
%% if required, the content of .bbl file can be included here once bbl is generated
%%\input sn-article.bbl

%% Default %%
%%\input sn-sample-bib.tex%

\end{document}